\documentclass[aps,showpacs,twocolumn,superscriptaddress,longbibliography]{revtex4-2} \usepackage{lipsum} 
\usepackage{physics}
\usepackage{amsmath} 
\usepackage{amsfonts} 
\usepackage{amssymb} 
\usepackage{epsfig} 
\usepackage{braket}
\usepackage{graphicx}
\usepackage{bbold} 
\usepackage{hyphenat} 
\usepackage{hyperref}  \usepackage{placeins}
\usepackage[dvipsnames]{xcolor}
\usepackage{array} 
\newcolumntype{C}[1]{>{\centering\arraybackslash}m{#1}}
\usepackage{overpic}

\usepackage[normalem]{ulem}
\usepackage{dcolumn} 
\usepackage{booktabs}
\usepackage{diagbox} 
\usepackage{siunitx} 
\usepackage{tabularx}

\AtBeginDocument{ \heavyrulewidth=.08em \lightrulewidth=.05em \cmidrulewidth=.03em \belowrulesep=.65ex \belowbottomsep=0pt \aboverulesep=.4ex \abovetopsep=0pt \cmidrulesep=\doublerulesep \cmidrulekern=.5em \defaultaddspace=.5em }
\begin{document}

\title{Collective mode spectroscopy in time-reversal
symmetry breaking superconductors}

\author{Silvia Neri}
\author{Walter Metzner}
\author{Dirk Manske}
\affiliation{ Max Planck Institute for Solid State Research, D-70569, Stuttgart, Germany}

\begin{abstract}
Collective excitations in superconductors provide essential insights into the symmetry of the broken phase, acting as indicators for identifying the ground state gap symmetry. Time-reversal symmetry breaking (TRSB) superconductors exhibit a rich spectrum of collective modes due to the complexity of their order parameters. These modes, known as “generalized clapping modes”, draw analogies to the clapping modes of Helium-3 Phase A. This study investigates two-dimensional TRSB superconductors with an order parameter of the form $\Delta = \Delta_1 + i\Delta_2$, exploring the characteristics of their collective mode spectrum. We begin with a phenomenological Ginzburg-Landau approach to build intuition, then develop a dynamical theory by deriving linearized equations of motion using the pseudospin formalism. Beyond the linear regime, we propose a classification scheme based on the potential to induce (an)isotropic oscillations in the superconducting condensate. By perturbing the system in symmetry channels distinct from the ground state, we aim to selectively enhance or suppress different mode responses. This study analyzes the features of these “generalized clapping modes” as a function of the ratio between the order parameter components under various excitation schemes. We believe that our findings could help distinguish between different order parameter symmetries in TRSB superconducting condensates and estimate the magnitude of their components.
\end{abstract}

\maketitle

\section{Introduction}

\begin{figure}[htpb!] 
    \centering 
\includegraphics[scale=.41]{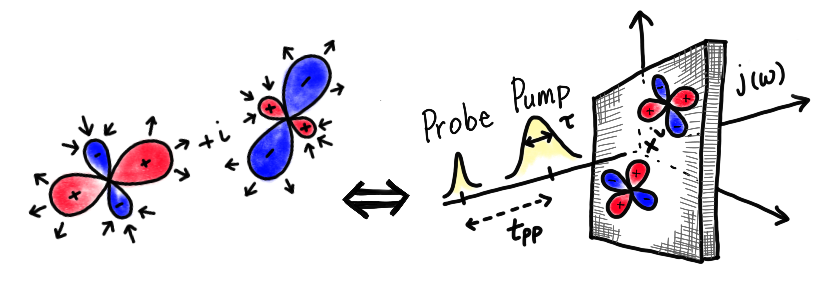} 
    \caption{Illustration of the core concept: by selectively inducing different types of oscillations in the condensate, we can excite the various modes characterizing the spectrum of collective excitations in a time-reversal symmetry-broken superconducting condensate. This process can be replicated using a laser pulse, with the excitation controlled by adjusting the in-plane polarization direction. } 
    \label{fig:conclusion} 
\end{figure}

The superconducting state is marked by the presence of various collective modes, such as Higgs, Leggett and Bardasis-Schrieffer mode among others \cite{PhysRevB.93.180507,Haenel_2021,PhysRevB.104.144508,M_ller_2019,PhysRevB.99.224510,PhysRevB.95.104503,Gabriele_2021,PhysRevB.90.224515,PhysRevB.104.174508}. The study and investigation of these modes can shed light on the superconducting ground state characteristics and on the symmetries of the order parameter. The emerging field of collective mode spectroscopy and,  particularly recent advancements in THz-range spectroscopic techniques \cite{Wu2024,Liu2025}, offers the opportunity to infer features on the dynamics of the order parameter by looking for fingerprints in the low energy spectrum of superconducting systems.\\
\indent One approach to excite the collective modes of the superconducting order parameter is to quench the system using an ultrafast, single-cycle THz pump pulse, driving the system out of equilibrium. This protocol was theoretically studied first in \cite{Amin_2004,PhysRevLett.96.230403,1974JETP8V} and later modeled with more realistic light pulses in \cite{PhysRevB.76.224522,PhysRevB.90.014515}.\\
\indent In this paper, we focus on an interesting category of unconventional superconductors that exhibit spontaneous time-reversal symmetry breaking (TRSB), with the aim of investigating the fingerprints of this symmetry breaking from the perspective of collective mode spectroscopy. \\ 
\indent The search for this class of exotic superconducting states can be traced back to the 1970s \cite{RevModPhys.47.331,PhysRev.123.1911,PhysRev.131.1553,Volovik1993,Rice_1995} with early studies on He$^3$, heavy fermions, Sr$_2$RuO$_4$, and then cuprates \cite{PhysRevLett.100.217004,PhysRevLett.100.127002} and iron-based superconductors \cite{Lee_2009,PhysRevLett.130.046702}. The proposals for their experimental detection have been mostly based on the possibility of detecting the internal magnetic field associated with the local alignment of the Cooper-pairs' magnetic moments as muon spin rotation and spontaneous Hall effect \cite{PhysRevLett.120.187004,condmat4020047}. \\
\indent Currently these TRSB superconductors have garnered significant attention due to their potential realization in various materials. Some examples are heavy-fermion superconductors, UPt$_3$ \cite{Schemm_2014}, URu$_2$Si$_2$ \cite{PhysRevB.91.140506}, UTe$_2$ \cite{Wei_2022}, some iron-based superconductors, as K-doped BaFe$_2$As$_2$, alongside engineered structures as twisted bilayers\cite{tbg} of TRS-preserving superconductors,  and Kagome superconductors \cite{Deng_2024} . \\
\indent In two-dimensional systems in particular breaking time-reversal provides a way for the system to stabilize to a fully gapped state while allowing pairing in multiple symmetry channels. The superconducting order parameter develops an internal structure such that the electrons pairing up in a Cooper-pair are in some relative motion one respect to the other. This results in a rich collective modes spectrum~\cite{PhysRevLett.84.4445,PhysRevB.100.094510} that shows modes similar in nature to those first theorized in the context of superfluid He$^3$ phase-A~\cite{Vollhardt1990,10.1093/acprof:oso/9780199564842.001.0001,Ling_1987}. These modes have been referred to in the literature as“generalized clapping modes”\cite{Narang,PhysRevResearch.6.043170,PhysRevB.109.094515}.\\  
\indent In particular, a two component order parameter is going to be characterized by a total of four bosonic excitations associated with the four degrees of freedom, two amplitude modes, and two phase modes that can in principle hybridize among each other. In the scenario that we are going to consider, the two components of the order parameter can either belong to a multi-dimensional irreducible representation or two different one-dimensional irreducible representations. In these cases, there is no symmetry allowed linear coupling between the components, and the phase and amplitude sectors decouple.\\
\indent In this work we propose a group theoretical classification of the possible oscillations that TRSB superconductors undergo when perturbed from the equilibrium state, extending the work in Ref. \cite{lukas} to include the case of even and odd parity two-component TRSB order parameters which are potentially relevant to experimental systems, these are: the d+id',  s+id and p+ip' -wave superconductors in two dimensions. To perturb the equilibrium state, we perform a quantum quench, akin to shrinking the Mexican hat potential, thereby driving the system out of equilibrium. We then investigate the dynamics within a pump-probe spectroscopic framework, applying a short, linearly polarized vector field to perturb the system. We study the full evolution of the gap components and present the numerical results obtained for both methods as a function of the ratio between the two components. The core idea of the paper is illustrated in Fig.~\ref{fig:conclusion}.\\
\indent  The paper is organized as follows: In Sec.~\ref{Sec:theo} we introduce the Ginzburg-Landau theory for a two component order parameter and the linearized pseudospin equations of motion. We also discuss the theoretical background of the two non-equilibrium protocols we selected to perturb the system with:  quantum quench and pump-probe. In Sec.~\ref{Sec:numerics} we describe the numerical methods adopted to solve the dynamics of the system. In particular, we use the pseudospin model to simulate the dynamics triggered by the quantum quenches and the density matrix formalism to investigate the motion in the presence of an external field. The latter method allows us to account for the transferred momentum of light to the system and then calculate the transient optical conductivity as a possible observable carrying the features of the mode structure of the superconducting state.  

\section{Theory and Methods}\label{Sec:theo}

We focus on time reversal symmetry breaking superconductors characterized by the order parameter below
\begin{equation}
    \Delta_{\boldsymbol{k}} = \Delta_{1,\boldsymbol{k}} \pm i \Delta_{2,\boldsymbol{k}}.
    \label{eq:OP}
\end{equation}
Here $\Delta_{i,\boldsymbol{k}}=\Delta_i f^i_{\boldsymbol{k}}$ with $i=1,2$, where $\Delta_{i}$ represents the magnitude of the superconducting order parameter and $f^i_{\boldsymbol{k}}$ the corresponding form factor. The two components can either belong to the same multi-dimensional irreducible representation or two irreducible representations. In particular, we consider this type of order parameter in the context of a clean superconductor with a single parabolic band that we model with a generalized BCS Hamiltonian at T=0. In Table \ref{tab:formfactor}, we specify the function used throughout the paper for the different form factors.
{\def\arraystretch{2}\tabcolsep=3 pt
\begin{table}[t!] 
    \centering
    \begin{tabular}{c|c|c}
    \toprule
        $f^i_{\boldsymbol{k}}$ (irrep) &Basis functions & Basis functions in polar coords.  \\\midrule
        $f^s_{\boldsymbol{k}}$ ($A_{1 g}$)  &   1     &   1\\
        $f^{d_{x^2-y^2}}_{\boldsymbol{k}}$ ($B_{1 g})$ & $x^2-y^2$ & $\cos (2 \varphi)$\\
        $f^{d_{xy}}_{\boldsymbol{k}}$ ($B_{2 g}$) & $x y$ & $\sin (2 \varphi)$ \\
        ($f^{p_{x}}_{\boldsymbol{k}},f^{p_{y}}_{\boldsymbol{k}}$) ($E_{u}$) & $(x,y)$ & $(\cos(\varphi),\sin (\varphi))$ \\
          \bottomrule
                
    \end{tabular}
    \caption{For each order parameter we consider in this work, we list the corresponding irreducible representation (irrep) of the tetragonal point group, the representative basis function in Cartesian coordinates, and as a function of the polar angle $\varphi$.}
    \label{tab:formfactor} 
\end{table}
}
To highlight the dependence of the mode structure on the ratio of the two components of the gap, we write it as:
\begin{equation}
    \Delta_{\boldsymbol{k}} = \Delta_0 (\eta_1 f^1_{\boldsymbol{k}} + i \eta_2 f^2_{\boldsymbol{k}}  ),~\quad
    \eta_1=\cos{\eta},~\eta_2 =\sin{\eta},
    \label{eq:eta}
\end{equation} 
such that $\tan{\eta}$ is the ratio between the two components. 

\subsection{Ginzburg-Landau Analysis}\label{Sec:GL}
We begin our investigation by analyzing the bosonic spectrum that follows from a general Ginzburg-Landau functional based on a free energy that respects the gauge and rotational symmetries of the system \cite{Garaud_2017, PhysRevB.53.2249, PhysRevB.56.11942}.
We start by considering a stationary free energy at fourth order
\begin{equation}
\begin{aligned}
\mathcal{F} & =\alpha_1 \left|\psi_1\right|^2+\alpha_2 \left|\psi_2\right|^2+\beta_1 \left|\psi_1\right|^4+\beta_2 \left|\psi_2\right|^4 \\
&+\beta_{12}\left|\psi_1\right|^2\left|\psi_2\right|^2+\beta_3\left(\left(\psi_1^*\right)^2 \psi_2^2+\left(\psi_2^*\right)^2 \psi_1^2\right).
\end{aligned} 
\label{eq:freeenergy}
\end{equation}
The collective excitations of the system correspond to fluctuations of the order parameter around the equilibrium configuration $\psi_{i,eq}$ (i=1,2) and $\theta_{2,eq}-\theta_{1,eq} = \pi/2$, therefore $\psi_{i,eq}=\Delta_0\eta_i$ in Eq.~\eqref{eq:eta}.
We then parameterize the order parameters as follows
\begin{equation}
\begin{aligned}
    &\psi_1 = (\psi_{1,eq}+h_1)e^{i\theta_1},\\
    &\psi_2 = (\psi_{2,eq}+h_2)e^{i(\pi/2 + \theta_2)},
\end{aligned}
\end{equation}
where $h_i, \theta_i$ are purely real and represent fluctuations around the ground state components $\psi_{i,eq}$.\\
Expanding the free energy in Eq.~\eqref{eq:freeenergy} to quadratic order in the small fluctuations $x = \{h_1,h_2,\theta_1,\theta_2\}$  we can compute the eigenmodes and their corresponding energies
\begin{equation}
\mathcal{F}=\mathcal{F}_0+\frac{1}{2} x_i M_{i j} x_j, \quad \text { where } \quad M=\left.\frac{\partial^2}{\partial x_i \partial x_j} \mathcal{F}\right|_{\boldsymbol{x}=0}.  \end{equation}
The matrix M has the eigenvectors :
\begin{equation}
\begin{aligned}
    u_+ &=( \frac{\beta_1 \psi_{1,eq}^2-\beta_2 \psi_{2,eq}^2 -b}{a},1,0,0),\\
    u_- &=(\frac{\beta_1 \psi_{1,eq}^2-\beta_2 \psi_{2,eq}^2 +b}{a},1,0,0),\\
    u_{\phi} &= (0,0,-1,1),\\
    u_{\textit{ABG}} &=(0,0,1,1),  
\end{aligned}
\label{eq:eigen}
\end{equation}
where we have defined
\begin{equation}
 \begin{aligned}
a=&\left(\psi_{1,eq} \psi_{2,eq} \right)^2\left(\beta_{12}-2\beta_3\right),\\
b=&[\beta_1^2 \psi_{1,eq}^4+\beta_2^2 \psi_{2,eq}^4 +(\beta^2_{12} - 2\beta_1\beta_2)\psi^2_{1,eq}\psi^2_{2,eq}\\
& + 4|\beta_3|(|\beta_3|- \beta_{12} )\psi^2_{1,eq}\psi^2_{2,eq}]^{1/2} ,
\end{aligned}  
\end{equation}
and the corresponding eigenvalues:

\begin{equation}
\begin{aligned}
&m_{+}=4(\beta_1 \psi_{1,eq}^2+\beta_2 \psi_{2,eq}^2 + b),\\
&m_{-}=4(\beta_1 \psi_{1,eq}^2+\beta_2 \psi_{2,eq}^2 - b),\\
&m_{\phi}=16 \beta_3 \psi^2_{1,eq}\psi^2_{2,eq}, \\
&m_{ABG}=0.
\end{aligned}     
\end{equation}

\begin{figure}[t!] 
    \centering 
\includegraphics[scale=.5]{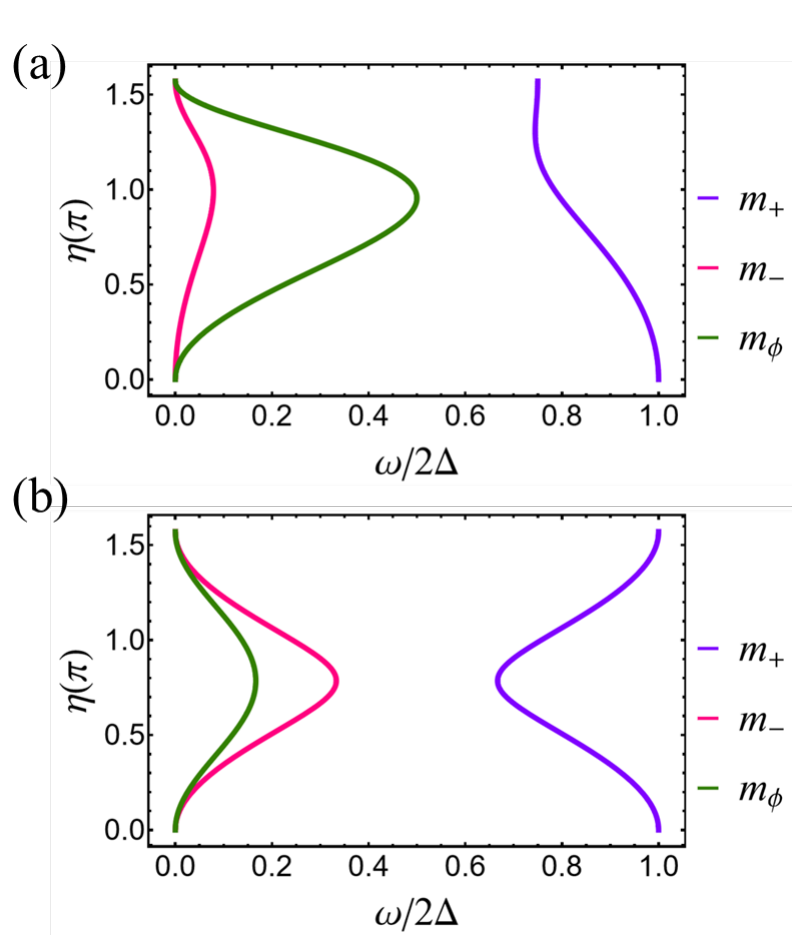} 
    \caption{  Collective modes energies normalized to the gap maxima, from the Ginzburg-Landau theory, as a function of mixing angle $\eta$   for the s+id in (a),  d+id', p+ip' case in (b), the latter two have the same spectra. It is possible to distinguish the global amplitude mode (or Higgs mode) $m_+$, the relative amplitude mode $m_-$, and the relative phase mode $m_{\phi}$. The Ginzburg-Landau coefficients used for the calculations are: in (a) $\beta_1=1$, $\beta_2=3/8$, $\beta_{12}=2$, $\beta_3=1/2$ (in agreement with \cite{PhysRevB.53.2249}); in (b) $\beta_1=3/2$, $\beta_2=3/2$, $\beta_{12}=2$, $\beta_3=1/2$ }. 
    \label{fig:gl} 
\end{figure}

The four collective bosonic modes are then a massless Anderson–Bogoliubov–Goldstone mode (ABG), a massive relative phase mode ($h_{\phi}$), and two massive amplitude modes ($h_+ , h_-$), which we are going to label as the Higgs mode and the relative amplitude mode in the following. 
We hereby obtain 
\begin{equation}
    \mathcal{F}=\mathcal{F}_0+\frac{1}{2} m_{-} h_{-}^2+\frac{1}{2} m_{+} h_{+}^2+\frac{1}{2} m_{\phi} h_{\phi}^2.
\end{equation}
Fig.~\ref{fig:gl}(a,b) displays the dispersion of the masses for the s+id, d+id', p+ip' case, where the latter two are degenerate.
As one could expect from the free energy functional in Eq.~\eqref{eq:freeenergy}, there are two modes corresponding to in-phase and out-of-phase oscillations of the order parameter amplitudes. Additionally, a mode arises from the phase fluctuations of $\theta_1$ and $\theta_2$, which is due to the quartic coupling term $\beta_3$. This term can be viewed as a second-order Josephson-like tunneling term and leads to the non-conservation of the particle number in each component. As a result, while we still have a Goldstone mode associated with the global U(1) symmetry, the relative phase mode acquires a finite mass. \\

\subsection{Linearized pseudospin formalism}\label{Sec:linpseudo}
 The Ginzburg-Landau framework does not take into account the quasiparticle excitation spectrum. To analyze the stability of the collective modes, we therefore need to perform a microscopic calculation.
We construct a theory for a superconducting state with a two-component order parameter on a single band. We consider a generalized BCS Hamiltonian with a gap of the form $\Delta_{{\boldsymbol{k}}} = \Delta_{1,{\boldsymbol{k}}} + i\Delta_{2,{\boldsymbol{k}}}$ coupled to a vector potential which represents a spatially homogeneous laser field. Here $\Delta_{i,k}=\Delta_i f^i_{\boldsymbol{k}}$, with $i=1,2$, where $f^{i}_{{\boldsymbol{k}}}$ is the form factor and, $f^1_{\boldsymbol{k}} ,f^2_{\boldsymbol{k}}$ are orthogonal. We consider then the two self consistency equations, 
\begin{equation}
\begin{aligned}
\Delta_1 &= -V_1\sum_{{\boldsymbol{k}}^{\prime}} \frac{ f^1_{{\boldsymbol{k}}^{\prime}}f^1_{{\boldsymbol{k}}^{\prime}} \Delta_1}{2 \sqrt{\epsilon^2_{{\boldsymbol{k}}^{\prime}} + \left|\Delta_{{\boldsymbol{k}}^{\prime}}\right|^2}}, \\
\Delta_2 &= -V_2\sum_{{\boldsymbol{k}}^{\prime}} \frac{ f^2_{{\boldsymbol{k}}^{\prime}}f^2_{{\boldsymbol{k}}^{\prime}} \Delta_2}{2 \sqrt{\epsilon^2_{{\boldsymbol{k}}^{\prime}} + \left|\Delta_{{\boldsymbol{k}}^{\prime}}\right|^2}},
\end{aligned}
\label{eq:bcs}
\end{equation}
where $\epsilon_{\boldsymbol{k}}$ is the band dispersion, $V_i$ represents the pairing interaction in the two channels, i=1,2, respectively, assuming the interaction to be a sum of two separable pairing interactions, as follows:
\begin{equation}
    V_{{\boldsymbol{k}},{\boldsymbol{k}}^{\prime}} = V_1 f^1_{{\boldsymbol{k}}} f^1_{{\boldsymbol{k}}^{\prime}} + V_2 f^2_{{\boldsymbol{k}}} f^2_{{\boldsymbol{k}}^{\prime}}.
    \label{eq:V}
\end{equation}
While writing the interaction, we already consider the case in which $f^1_{{\boldsymbol{k}}}$ and $f^2_{{\boldsymbol{k}}}$ are orthogonal to each other, such that there is no mixing between them.\\
\indent Following the steps outlined in \cite{PhysRevB.95.104503}, we introduce Anderson's pseudospin \cite{PhysRev.112.1900},
\begin{equation}
\boldsymbol{\sigma}_{{\boldsymbol{k}}}=\frac{1}{2} \left(\begin{array}{c}
c_{{\boldsymbol{k}} \uparrow}^{\dagger},~c_{-{\boldsymbol{k}} \downarrow}
\end{array}\right) \boldsymbol{\tau}  \left(\begin{array}{c}
c_{{\boldsymbol{k}} \uparrow} \\
c_{-{\boldsymbol{k}} \downarrow}^{\dagger}
\end{array}\right),
\label{eq:pseudo}
\end{equation}
with the Pauli matrices $\tau_i$. 
The equilibrium expectation values for the pseudospins at T=0 read
\begin{equation}
   \begin{aligned}
\left\langle\sigma_{\boldsymbol{k}}^x\right\rangle^{eq} & =\frac{\Delta_{\boldsymbol{k}}^{\prime}}{2 E_{\boldsymbol{k}}} \\
\left\langle\sigma_{\boldsymbol{k}}^y\right\rangle^{eq} & =-\frac{\Delta_{\boldsymbol{k}}^{\prime \prime}}{2 E_{\boldsymbol{k}}}  \\
\left\langle\sigma_{\boldsymbol{k}}^z\right\rangle^{eq} & =-\frac{\epsilon_{\boldsymbol{k}}}{2 E_{\boldsymbol{k}}}
\end{aligned} 
\end{equation}\label{eq:bloch_eq}

\begin{figure}[t!] 
    \centering 
\includegraphics[scale=.9]{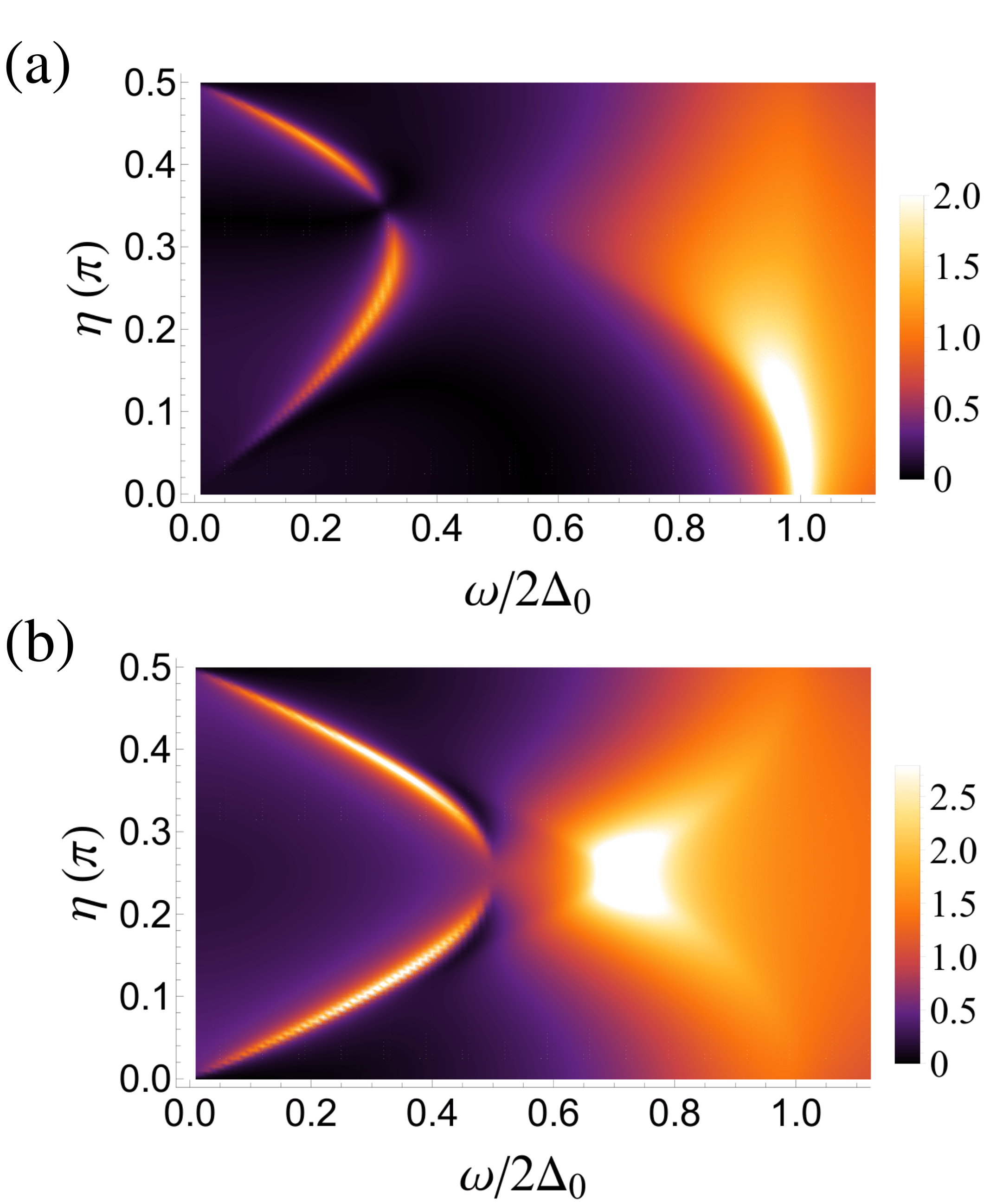} 
    \caption{ Collective modes energies as a function of mixing angle $\eta$ for the s+id in (a),  d+id', p+ip' case in (b), the latter two have the same spectra. Fourier spectrum (in arbitrary units) $|\Delta(\omega)|=|FT|\Delta(t)||$, obtained from the linearized equations of motion in Eq.~\eqref{eq:linear}, as a function of frequency $\omega$ (scaled to the gap at equilibrium $2\Delta_0$) and the mixing angle $\eta$. Results obtained by assuming $c_1 \neq 0$ to unveil the structure of the amplitude sector (see main text for further details)}. 
    \label{fig:linear} 
\end{figure}

\begin{figure}[t!] 
    \centering 
\includegraphics[scale=.7]{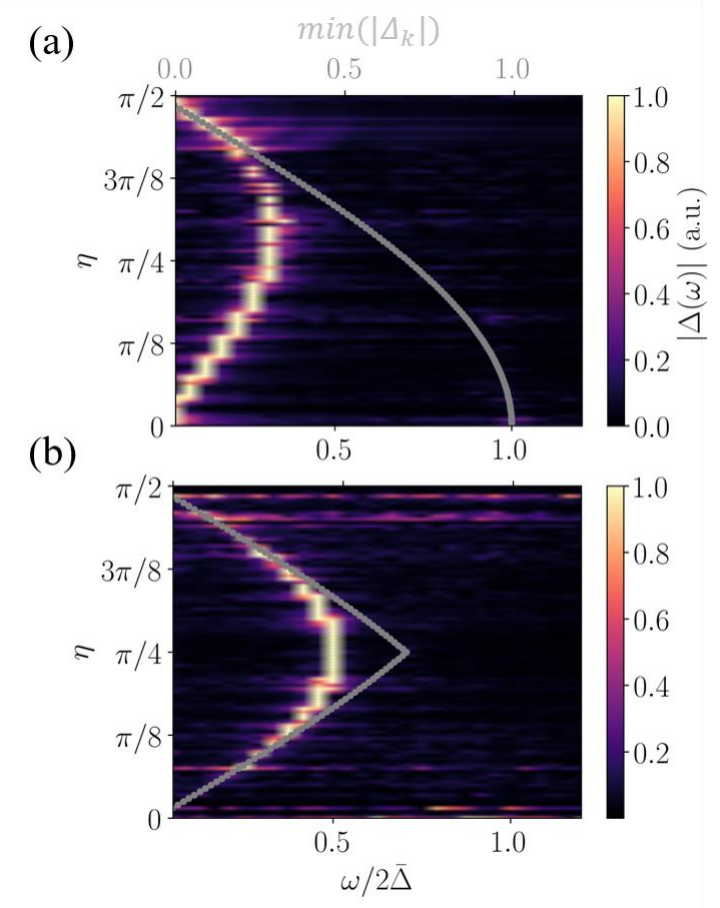} 
    \caption{Relative phase mode as a function of mixing angle $\eta$ for the s+id in (a),  d+id', p+ip' case in (b), the latter two have the same spectra. The plot represents the Fourier spectrum (in arbitrary units) $|\Delta(\omega)|=|FT|\Delta(t)||$, as a function of frequency $\omega$ (scaled to the asymptotic value of the gap $2\bar{\Delta}$) and the mixing angle $\eta$. The relative phase mode branches out from the quasiparticle continuum $min(|\Delta_k|)$, gray line in the plots. The calculations are performed within the full pseudospin formalism by numerically solving Eq.~\eqref{eq:bloch}. We consider a single-cycle THz pulse with the Gaussian-shaped envelope ($A_0=0.1$,~$\hbar\Omega=0.4~\mathrm{meV}$,~$\tau=4~\mathrm{ps}$). The parameters used in the calculations are:  $\epsilon_{{\boldsymbol{k}}}=t_{hop}{{\boldsymbol{k}}}^2-\epsilon_F$, with parameters $t_{hop}= 50$~meV for the nearest-neighbor hopping and Fermi energy $\epsilon_F$= 100~meV. The equilibrium gap value is chosen to be $|\Delta_{0}|$= 1~meV. The grid is chosen in polar coordinates, with $\Delta\varphi$ spacing in azimuthal direction and radial spacing $\Delta\epsilon$, the lower and upper bound of the energy cutoff is $\epsilon_{\boldsymbol{k}}=\pm\epsilon_c$ with $\epsilon_c=10$~meV, with number of points $N_\epsilon=2000$, $N_\varphi=101$ for $t_{max}=30$~ps.} 
    \label{fig:pumpprobe_pseudo} 
\end{figure}

The equations of motion of the Anderson pseudospins with respect to the BCS Hamiltonian coupled to the vector potential describe then a precession around the pseudomagnetic field $\mathbf{b}_{{\boldsymbol{k}}}$\cite{PhysRevB.92.064508}
\begin{equation}
 \partial_t \boldsymbol{\sigma}_{{\boldsymbol{k}}}(t)=\mathrm{i}\left[H_{\mathrm{BCS}}(t), \boldsymbol{\sigma}_{{\boldsymbol{k}}}(t)\right]=\mathbf{b}_{{\boldsymbol{k}}}(t) \times \boldsymbol{\sigma}_{{\boldsymbol{k}}}(t) ,  
 \label{eq:bloch}
\end{equation}
where $H_{BCS}$ represents the BCS Hamiltonian coupled to a vector potential $\boldsymbol{A}(t)$. We work in a gauge in which the scalar potential is set to zero. The vector potential enters the Hamiltonian via minimal coupling ${\boldsymbol{k}} \rightarrow {\boldsymbol{k}}-e \boldsymbol{A}(t)$ with the electron charge e.
\begin{equation}
    \begin{aligned}
\mathcal{H}_{BCS}(t)= & \sum_{{\boldsymbol{k}} \sigma} \epsilon_{({\boldsymbol{k}}-e \boldsymbol{A}(t))} c_{{\boldsymbol{k}} \sigma}^{\dagger} c_{{\boldsymbol{k}} \sigma}- \sum_{{\boldsymbol{k}}} \Delta_{{\boldsymbol{k}}}c_{{\boldsymbol{k}} \uparrow}^{\dagger} c_{-{\boldsymbol{k}} \downarrow}^{\dagger} \\
& - \sum_{{\boldsymbol{k}}} \Delta_{{\boldsymbol{k}}}^* c_{-{\boldsymbol{k}} \downarrow} c_{{\boldsymbol{k}} \uparrow}
\end{aligned}
\end{equation}
expressing the same Hamiltonian within the pseudospin formalism, we define $\boldsymbol{b_{\boldsymbol{k}}}$ as follows
\begin{equation}
\mathbf{b}_{{\boldsymbol{k}}}=\left(-2 \Delta_{\boldsymbol{k}}^{\prime} , 2 \Delta_{\boldsymbol{k}}^{\prime\prime},\epsilon_{{\boldsymbol{k}}-e\mathbf{A(t)}} + \epsilon_{{\boldsymbol{k}}+e\mathbf{A(t)}}  \right).    
\end{equation}
We then linearize the Bloch equations for deviations from equilibrium within the approximation of a small laser intensity
\begin{equation}
  \begin{aligned}
\left\langle\sigma_{{\boldsymbol{k}}}^x\right\rangle(t) & =\left\langle\sigma_{{\boldsymbol{k}}}^x\right\rangle^{\mathrm{eq}}+\delta\sigma_{{\boldsymbol{k}}}^x(t), \\
\left\langle\sigma_{{\boldsymbol{k}}}^y\right\rangle(t) & =\left\langle\sigma_{{\boldsymbol{k}}}^y\right\rangle^{\mathrm{eq}}+\delta\sigma_{{\boldsymbol{k}}}^y(t), \\
\left\langle\sigma_{{\boldsymbol{k}}}^z\right\rangle(t) & =\left\langle\sigma_{{\boldsymbol{k}}}^y\right\rangle^{\mathrm{eq}}+\delta\sigma_{{\boldsymbol{k}}}^z(t), \\
\Delta_{{\boldsymbol{k}}}(t) & = \Delta_{1,eq}f^1_{{\boldsymbol{k}}}+i\Delta_{2,eq}f^2_{{\boldsymbol{k}}}+ \delta \Delta_1(t)f^1_{{\boldsymbol{k}}}+\delta \Delta_2(t)f^2_{{\boldsymbol{k}}} ,
\end{aligned}\label{eq:linear}  
\end{equation}
where the fluctuations, $\delta\Delta_i$, are complex numbers.\\
The effect of the laser field enters the equations via $\delta b^{z}_{{\boldsymbol{k}}}(t) = b^{z}_{{\boldsymbol{k}}}(t) - \epsilon_{{\boldsymbol{k}}} \approx \frac{e^2}{2} \sum_{ij}(\partial_{k_i}\partial_{k_j} \epsilon_{\boldsymbol{k}} ) A_i(t)A_j(t) $.\\
From the linearized equations of motion, we derive solutions for the real and imaginary components of the gap in the form
 \begin{equation}
\left(\begin{array}{c}
\delta\Delta^{'}_{{\boldsymbol{k}},1} \\
\delta\Delta^{''}_{{\boldsymbol{k}},1}\\
\delta\Delta^{'}_{{\boldsymbol{k}},2}\\
\delta\Delta^{''}_{{\boldsymbol{k}},2}
\end{array}\right) \propto e^2 A^2(\omega)\left(\begin{array}{c}
\Delta^{'}_{{\boldsymbol{k}},1} \\
\Delta^{''}_{{\boldsymbol{k}},1}\\
\Delta^{'}_{{\boldsymbol{k}},2}\\
\Delta^{''}_{{\boldsymbol{k}},2}
\end{array}\right).
 \end{equation}
 From this analysis in particular, see Appendix~\ref{app:linearpseudo} for further details, we are able to infer the coupling between these modes and the vector potential. While the amplitude sector couples via $\left(e^2 A^2 / 2\right) \left(\partial^2_{k_x} \epsilon_{{\boldsymbol{k}}}\right) \epsilon_{{\boldsymbol{k}}} $, as it is the case for the Higgs mode \cite{PhysRevB.96.014503,PhysRevB.93.180507},  the phase fluctuations couple via $\left(e^2 A^2 / 2\right) \left(\partial^2_{k_x} \epsilon_{{\boldsymbol{k}}}\right)$,  as it is the case for the Leggett mode \cite{PhysRevB.95.104503,PhysRevB.94.064512} and Bardasis-Schrieffer mode \cite{rafael,PhysRevB.104.144508}.
We can then derive an equation for the relative phase mode as follows: since in the linearized equations the imaginary fluctuations $\delta\Delta^{"}_1$ of the real component of the gap are proportional to the phase $\theta_{1}$ and the real fluctuations $\delta\Delta_2$ of the second component is proportional to the phase $\theta_2$; we can compute the equation for the phase difference between the two components of the gap, which is a gauge-invariant quantity, as :
\begin{equation}
\begin{aligned}
        \delta[ \theta_1(\omega)-\theta_2(\omega)]=\frac{\delta\Delta_1^{"}}{\Delta_{1,eq}} +\frac{\delta\Delta_2^{'}}{\Delta_{2,eq}}=\frac{e^2 A(\omega)^2}{2} \partial_{\boldsymbol{k}}^2 \epsilon_{\boldsymbol{k}}  i\omega \textit{L}(\omega)\\= \frac{e^2 A(\omega)^2}{2} c_0  i\omega \textit{L}(\omega) ,
\end{aligned}
\end{equation}
where $L(\omega)$ is the propagator of the relative phase mode. Here we used the fact that for generic band structures, one can write  \cite{PhysRevB.95.104503} 
\begin{equation}
\sum_{{\boldsymbol{k}}} \delta\left(\epsilon-\epsilon_{{\boldsymbol{k}}}\right) \frac{\partial^2 \epsilon_{{\boldsymbol{k}}}}{\partial k_i^2}=D\left(c_{0}+c_{1} \epsilon+c_{2} \epsilon^2 \cdots\right),    
\end{equation}\label{eq:banddisperion}
where i=x,y.\\
\indent By solving the equations in Appendix~\ref{app:linearpseudo}, we obtain $\textit{L}(\omega)=0$ . In agreement with the results in Ref.~\cite{PhysRevB.92.094506}, the relative phase mode does not appear within the linear regime.\\
\indent We can, furthermore, derive the associated equations for the relative amplitude mode, defined in the linearized regime, as $ \frac{\delta\Delta_1^{"}}{\Delta_{1,eq}} -\frac{\delta\Delta_2^{'}}{\Delta_{2,eq}}$
and for the Higgs mode, or global amplitude mode, $ \frac{\delta\Delta_1^{'}}{\Delta_{1,eq}}+\frac{\delta\Delta_2^{"}}{\Delta_{2,eq}}$.
From Eq.~\eqref{eq:fluct} in Appendix~\ref{app:linearpseudo} is possible to see that these two modes associated with the amplitude sector couple to light as $\propto \partial_{\boldsymbol{k}}^2 \epsilon_{\boldsymbol{k}}~\epsilon_{\boldsymbol{k}}$ or, equivalently $\propto c_1$. This means that in a clean superconductor with a parabolic band dispersion, these amplitude modes do not couple with light. \\
To understand the structure of the modes with respect to the ratio between the two components of the gap, we exploit the two results obtained:
\begin{itemize}
    \item Within the linear regime, the relative phase mode is identically zero.
    \item The amplitude sector is the only one present, but since for a parabolic band dispersion, $c_1=0$, it does not couple to light.
\end{itemize}
We then set $c_1 = 1$, and we plot the results in Fig.~\ref{fig:linear}. This “fictitious” substitution allows us to unveil the structure of the propagators. The plot then only contains the amplitude sector.\\ 
To obtain the dynamics of the phase mode, we instead proceed by numerically solving the full equation of motion for the Anderson pseudospins in Eq.~\eqref{eq:bloch}. The results for the s+id and d+id' case are shown in Fig.~\ref{fig:pumpprobe_pseudo}. Here, since in the numerical simulation the band dispersion is parabolic, the amplitude sector naturally does not appear. To model A(t), we chose a single-cycle THz pulse with a Gaussian envelope $A(t)={A}_0 \mathrm{e}^{-4 \ln (2)\left(\frac{t}{\tau}\right)^2} \cos \left(\Omega t \right) $.
Comparing the results shown in Fig.~\ref{fig:linear} with Fig.~\ref{fig:gl}, we can clearly identify the  Higgs mode $m_+$ and the relative amplitude mode $m_-$.  Specifically, in the $s+id$ case, the mode we refer to here as the relative amplitude mode has previously been identified in the literature as the mixed-symmetry Bardasis-Schrieffer mode \cite{mbs}. \\

\subsection{Quantum quenches}\label{Sec:qq}
\begin{figure}[t!] 
    \centering 
 \includegraphics[scale=0.425]{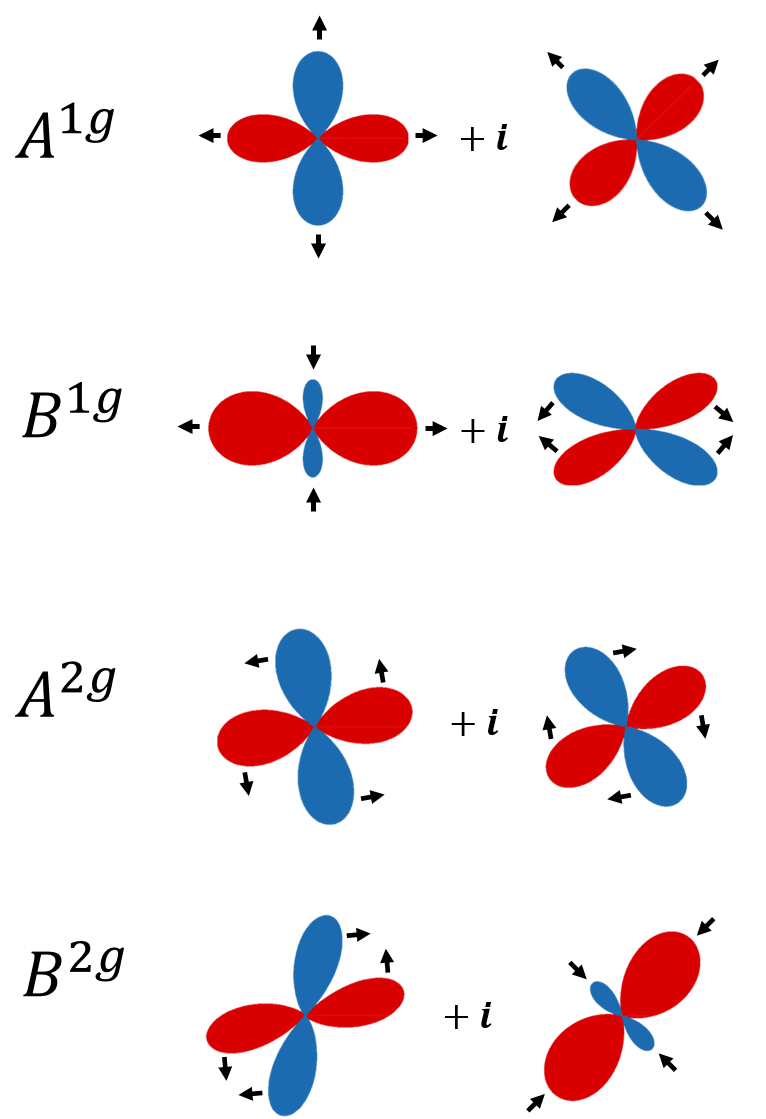} 
    \caption{Possible even oscillation symmetries for $d_{x^2 - y^2}~+~id_{xy}$ superconductor with point group symmetry $D_{4h}$ of the underlying lattice. The arrows represent the motion of the lobes over time.} 
    \label{fig:did_osc}
\end{figure}

{\def\arraystretch{2}\tabcolsep=2 pt
\begin{table}[htpb!] 
    \centering
    \begin{tabular}{c|c|c}
    \toprule
        Gap &Quench symmetry $f^q_{\boldsymbol{k}}$& Condensate osc.  \\\midrule
        $s + id_{x^2 - y^2}$  &   $1 + i~(x^2 - y^2)$     &   $A^{1g}_{s+id}$\\
                &  $ x^2 - y^2 + i~1$     &   $B^{1g}_{s+id}+A^{1g}_{s+id}$\\
                &   $xy(x^2 - y^2) + i~xy$  &   $A^{2g}_{s+id}+A^{1g}_{s+id}$\\
                &   $xy + i~xy(x^2 - y^2)$  &   $B^{2g}_{s+id}+A^{1g}_{s+id}$\\ \midrule
        $d_{x^2 - y^2} + id_{xy}$ &  $x^2 - y^2 +i~xy$     &   $A^{1g}_{d+id'}$\\
                &   $1 + i~xy(x^2 - y^2) $  &   $B^{1g}_{d+id'}+A^{1g}_{d+id'}$\\
                &   $xy + i~(x^2 - y^2) $   &   $A^{2g}_{d+id'}+A^{1g}_{d+id'}$\\
                &   $xy(x^2 - y^2) + i~1 $   &   $B^{2g}_{d+id'}+A^{1g}_{d+id'}$\\ \midrule
                $p_x + i p_y$ &  $x +i~y$     &   $A^{1g}_{p+ip'}$\\
                &   $x(x^2 - y^2) + i~y(x^2 - y^2) $  &   $B^{1g}_{p+ip'}+A^{1g}_{p+ip'}$\\
                &   $x^2y + i~xy^2 $   &  
                $B^{2g}_{p+ip'}+A^{1g}_{p+ip'}$\\
                &   $x^2y(x^2 - y^2) + i~xy^2(x^2 - y^2) $   &   $A^{2g}_{p+ip'}+A^{1g}_{p+ip'}$\\ \bottomrule
                
    \end{tabular}
    \caption{Summary of condensate oscillations and corresponding quench symmetries.
    The first column depicts the considered gap symmetries. The second column lists the momentum dependent quenches in the case of a $D_{4h}$ point group. The third column lists the induced condensate oscillations. We present the quench symmetry adopted to perturb the $s+id_{x^2 - y^2}$, $d_{x^2 - y^2} + i d_{xy}$ and $p_x + ip_y$ case. In the latter, in order to trigger even oscillation of the order parameter, we adopted quench functions that are just the even basis functions of the point group multiplied by the doublet [x,y].\\}
    \label{tab:quenches} 
\end{table}
}
To proceed further in the investigation of the structure of the collective modes of these superconducting systems with our minimal model of a clean BCS superconductor with a parabolic band dispersion at T=0, we here adopt another excitation scheme: we perform quantum quenches on the superconducting state to excite the modes.\\
\indent The goal here is to induce some dynamics by "deforming" the ground state symmetry to trigger oscillations of the superconducting condensate in symmetry channels different than the ground state symmetry, see Fig.~\ref{fig:did_osc} for a pictorial representation of the possible dynamics we can start for the case of a $d_{x^2 - y^2}+i d_{xy}$ superconductor. 
To do so, we numerically solve the Bloch equations $ \partial_t \boldsymbol{\sigma}_{{\boldsymbol{k}}}(t)=\mathbf{b}_{{\boldsymbol{k}}}(t) \times \boldsymbol{\sigma}_{{\boldsymbol{k}}}(t) $ for the pseudospins Eq.~\eqref{eq:pseudo}, but, instead of coupling the pseudospins to a vector potential, we perturb the system with a momentum dependent quench, see Ref.~\cite{lukas}. Here $\mathbf{b}_{{\boldsymbol{k}}}=\left(-2 \Delta_{\boldsymbol{k}}^{\prime} , 2 \Delta_{\boldsymbol{k}}^{\prime\prime},2\epsilon_{{\boldsymbol{k}}}  \right).$
Before the quench we have for the $\sigma_{x}$ and $\sigma_{y}$:
\begin{equation}
\langle \sigma_{{\boldsymbol{k}}}^{x} \rangle^{eq} = \frac{\Delta_1 f^1_{{\boldsymbol{k}}}}{2 E_{{\boldsymbol{k}}}} \ , \quad \langle \sigma_{{\boldsymbol{k}}}^{y} \rangle^{eq} = -\frac{\Delta_2 f^2_{{\boldsymbol{k}}}}{2 E_{{\boldsymbol{k}}}} \ .
\end{equation}
At t=0, we apply a state quench where we modify the symmetry of the condensate by changing the pseudospin expectation values to;
\begin{equation}
\langle \sigma_{{\boldsymbol{k}}}^{x} \rangle(0) = \frac{\Delta_1 \bar{f}^1_{{\boldsymbol{k}}}}{2 \bar{E}_{{\boldsymbol{k}}}} \ , \quad \langle \sigma_{{\boldsymbol{k}}}^{y}\rangle(0) =- \frac{\Delta_2 \bar{f}^2_{{\boldsymbol{k}}}}{2 \bar{E}_{{\boldsymbol{k}}}} \ 
\end{equation}
with $\bar{f}_{{\boldsymbol{k}}}^{(i=1,2)}=f^{(i=1,2)}_{{\boldsymbol{k}}}+\delta f_{{\boldsymbol{k}}}^{\mathrm{q}(i=1,2)}$, where $f_{{\boldsymbol{k}}}^{(i=1,2)}$ is the form factor of each component at equilibrium and $f_{{\boldsymbol{k}}}^{\mathrm{q}(i=1,2)}$ is the quench form factor for each component, see Tab.~\ref{tab:quenches}, $\delta$ is the quench strength. The quench alters the quasiparticle distribution of the condensate such that $\bar{E}_{{\boldsymbol{k}}}=\sqrt{\epsilon^2_{{\boldsymbol{k}}}+|\Delta_{1} \bar{f}^{1}_{{\boldsymbol{k}}}|^2+|\Delta_{2} \bar{f}^{2}_{{\boldsymbol{k}}}|^2}$, this can lead to the dynamical appearance of additional modes for certain quench symmetries, see Ref.~\cite{lukas}.
Such a sudden change drives the system out of equilibrium~\cite{Yuzbashyan_2005,Yuzbashyan_2006,PhysRevLett.115.257001,PhysRevB.78.132505}.\\
Within this setting, we aim to investigate whether it is possible to enhance the response of certain excitations by triggering oscillations of the system in symmetry channels different from the ground state symmetry.\\
In this scenario, the coherent response of the superconducting condensate to the perturbation is pinned to the symmetry of the lattice. We consider a lattice with $D_{4h}$ space symmetry; therefore, there are four different irreducible representations: $A^{1g}$, $A^{2g}$,$ B^{1g}$, $B^{2g}$. The condensate oscillations can be decomposed into contributions from these symmetry sectors as in Fig.~\ref{fig:did_osc}.
\\
\indent We list in Tab.~\ref{tab:quenches} the quantum quenches that we are going to apply in Sec.~\ref{Sec:numerics_qq}, along with the corresponding oscillations they trigger, for the \( s + id_{x^2 - y^2} \), \( d_{x^2 - y^2} + id_{xy} \), and \( p_x + i p_y \) cases. 
For the three cases presented in column one, we list in column two the momentum dependent quench function and in column three the symmetry of the condensate oscillation. As it appears, the $A^{1g}$ isotropic oscillation is always excited. 

\subsection{Pump-probe }\label{Sec:pp}
\begin{figure}[htpb!]
    \centering 
\includegraphics[scale=.56]{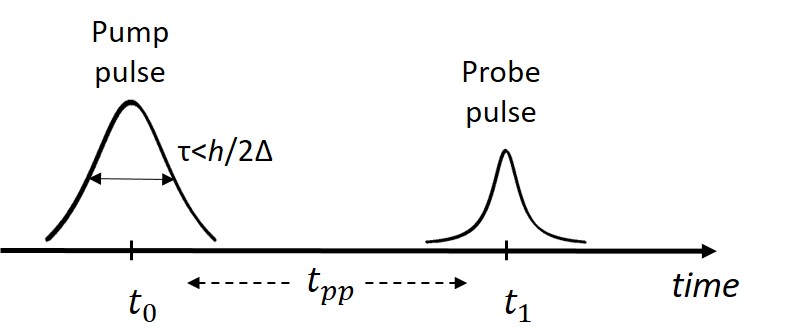}
    \caption{Schematic picture of a pump-probe experiment. The pump pulse is applied at time $t_0$ and
acts as a quench to trigger the non-equilibrium dynamics. After a variable time delay $t_{pp}$, a second weak
probe pulse is applied to measure the instantaneous state of the system.}
    \label{fig:pumpprobe}
\end{figure}

In this section, we present the last among the various schemes adopted in the present paper to investigate the dynamics of the “generalized clapping modes”. In particular, this latter scheme aims to replicate the dynamics triggered by the momentum-dependent quantum quenches introduced in the previous section, within a pump-probe excitation scheme. \\
Pump-probe spectroscopy has been established as a useful tool to probe the dynamical properties of superconductors \cite{Dal_Conte_2012,PhysRevLett.105.067001}, a schematic of the technique is shown in Fig.~\ref{fig:pumpprobe}. First, an intense laser pulse, the pump, drives the system to non-equilibrium. After a controlled delay, a second laser pulse, the probe, is used to measure the response to the pump pulse. By varying the delay time between the probe and the pump, it is possible to capture the dynamics of the superconducting state.\\
We want to remind here, however, that in Sec.~\ref{Sec:linpseudo}, we underlined that by adopting the pseudospin formalism, we derived that the form of the coupling between the amplitude sector and a light pulse is such that, within our model, the amplitude sector does not couple to light. \\
We show here that, going beyond the pseudospin formalism and retaining information about the momentum transferred by the light pulse, which we need in order to mimic the above-mentioned momentum-dependent quenches, the picture changes. 
To theoretically study the response of our systems in the spectroscopic setup described, we adopt the density matrix formalism \cite{RevModPhys.74.895,papenkort}. 
To solve the dynamics, we use the method outlined in Ref.~\cite{papenkort} based on an expansion of Heisenberg's equations of motion. For this, it is advantageous to perform a Bogoliubov transformation of the BCS Hamiltonian and calculate the dynamics of expectation values of the Bogoliubov operators,
\begin{equation}
\alpha_{{\boldsymbol{k}}}=u_{{\boldsymbol{k}}} c_{{\boldsymbol{k}}\uparrow}-v_{{\boldsymbol{k}}} c_{-{\boldsymbol{k}} \downarrow}^{\dagger}, \quad \beta_{{\boldsymbol{k}}}=v_{{\boldsymbol{k}}} c^{\dagger}_{{\boldsymbol{k}}\uparrow}+u_{{\boldsymbol{k}}} c_{-{\boldsymbol{k}} \downarrow},    
\end{equation}
with $u_{{\boldsymbol{k}}}=\sqrt{\left(1+\epsilon_{{\boldsymbol{k}}} / E_{{\boldsymbol{k}}}\right) / 2}$ and $v_{{\boldsymbol{k}}}=\sqrt{\left(1-\epsilon_{{\boldsymbol{k}}} / E_{{\boldsymbol{k}}}\right) / 2}$. 
All physical observables, such as the order parameter amplitude $\left|\Delta_{{\boldsymbol{k}}}(t)\right|$ can now be expressed in terms of the Bogoliubov quasiparticle expectation values $\langle\alpha_{{\boldsymbol{k}}}^{\dagger} \alpha_{{\boldsymbol{k}}^{\prime}}\rangle$, $\langle\beta_{{\boldsymbol{k}}}^{\dagger} \beta_{{\boldsymbol{k}}^{\prime}}\rangle,\langle\alpha_{{\boldsymbol{k}}}^{\dagger} \beta_{{\boldsymbol{k}}^{\prime}}^{\dagger}\rangle$ and $\left\langle\alpha_{{\boldsymbol{k}}} \beta_{{\boldsymbol{k}}^{\prime}}\right\rangle$. For each expectation value, we evaluate the commutator, for example
\begin{equation}
 \partial_t \langle\alpha_{{\boldsymbol{k}}}^{\dagger} \alpha_{{\boldsymbol{k}}^{\prime}}\rangle(t)=\mathrm{i}\left\langle\left[H(t), \alpha_{{\boldsymbol{k}}}^{\dagger} \alpha_{{\boldsymbol{k}}^{\prime}}\right]\right\rangle(t).
 \label{eq:bloch_qp}
\end{equation}
Here H represents the BCS Hamiltonian coupled to a vector potential, the expression is derived in Appendix~\ref{app:eom} in terms of  the fermionic operators, which after the transformation reads
\begin{equation}
    \begin{aligned}
H&=H_{BCS}+H^{(1)}_{em}+H^{(2)}_{em},\\
H_{BCS} & =\sum_{{\boldsymbol{k}}}\left(\begin{array}{cc}
\alpha_{{\boldsymbol{k}}}^{\dagger}, & \beta_{{\boldsymbol{k}}}
\end{array}\right)\left(\begin{array}{cc}
R_{{\boldsymbol{k}}} & C_{{\boldsymbol{k}}} \\
C_{{\boldsymbol{k}}}^* & -R_{{\boldsymbol{k}}}
\end{array}\right)\binom{\alpha_{{\boldsymbol{k}}}}{\beta_{{\boldsymbol{k}}}^{\dagger}} \\
& =\sum_{{\boldsymbol{k}}} R_{{\boldsymbol{k}}}\left(\alpha_{{\boldsymbol{k}}}^{\dagger} \alpha_{{\boldsymbol{k}}}+\beta_{{\boldsymbol{k}}}^{\dagger} \beta_{{\boldsymbol{k}}}\right)+\sum_{{\boldsymbol{k}}} C_{{\boldsymbol{k}}} \alpha_{\boldsymbol{k}}^{\dagger} \beta_{\boldsymbol{k}}^{\dagger}-C_{\boldsymbol{k}}^* \alpha_{\boldsymbol{k}} \beta_{\boldsymbol{k}},\\
H_{\mathrm{em}}^{(1)}= & \frac{e \hbar}{2 m} \sum_{{\boldsymbol{k}}, \boldsymbol{q}}(2 {\boldsymbol{k}}+\boldsymbol{q}) \boldsymbol{A}_{\boldsymbol{q}}(t)\left(L_{{\boldsymbol{k}}, \boldsymbol{q}}^{(+)} \alpha_{{\boldsymbol{k}}+\boldsymbol{q}}^{\dagger} \alpha_{{\boldsymbol{k}}}-L_{{\boldsymbol{k}}, \boldsymbol{q}}^{(+) *} \beta_{{\boldsymbol{k}}}^{\dagger} \beta_{{\boldsymbol{k}}+\boldsymbol{q}}\right. \\
& \left.\quad+M_{{\boldsymbol{k}}, \boldsymbol{q}}^{(-) *} \alpha_{{\boldsymbol{k}}+\boldsymbol{q}}^{\dagger} \beta_{{\boldsymbol{k}}}^{\dagger}+M_{{\boldsymbol{k}}, \boldsymbol{q}}^{(-)} \alpha_{{\boldsymbol{k}}} \beta_{{\boldsymbol{k}}+\boldsymbol{q}}\right), \\
H_{\mathrm{em}}^{(2)}= & \frac{e^2}{2 m} \sum_{{\boldsymbol{k}}, \boldsymbol{q}}\left[\sum_{\boldsymbol{q}^{\prime}} \boldsymbol{A}_{\boldsymbol{q}-\boldsymbol{q}^{\prime}}(t) \cdot \boldsymbol{A}_{\boldsymbol{q}^{\prime}}(t)\right]\left(L_{{\boldsymbol{k}}, \boldsymbol{q}}^{(-)} \alpha_{{\boldsymbol{k}}+\boldsymbol{q}}^{\dagger} \alpha_{{\boldsymbol{k}}}\right.\\
& \left.+L_{{\boldsymbol{k}}, \boldsymbol{q}}^{(-) *} \beta_{{\boldsymbol{k}}}^{\dagger} \beta_{{\boldsymbol{k}}+\boldsymbol{q}} 
+M_{{\boldsymbol{k}}, \boldsymbol{q}}^{(+) *} \alpha_{{\boldsymbol{k}}+\boldsymbol{q}}^{\dagger} \beta_{{\boldsymbol{k}}}^{\dagger}-M_{{\boldsymbol{k}}, \boldsymbol{q}}^{(+)} \alpha_{{\boldsymbol{k}}} \beta_{{\boldsymbol{k}}+\boldsymbol{q}}\right),
\end{aligned}
\end{equation}
with the following abbreviations
\begin{equation}
\begin{aligned}
& R_{{\boldsymbol{k}}}(t)=\epsilon_{{\boldsymbol{k}}}\left(\left|u_{\boldsymbol{k}}\right|^2-\left|v_{\boldsymbol{k}}\right|^2\right)+\Delta_{\boldsymbol{k}}(t) u_{\boldsymbol{k}} v_{\boldsymbol{k}}^*+\Delta_{\boldsymbol{k}}^*(t) u_{\boldsymbol{k}}^* v_{\boldsymbol{k}}, \\
& C_{\boldsymbol{k}}(t)=2 \epsilon_{\boldsymbol{k}} u_{\boldsymbol{k}} v_{\boldsymbol{k}}-\Delta_{\boldsymbol{k}}(t) u_{\boldsymbol{k}}^2+\Delta_{\boldsymbol{k}}^*(t) v_{\boldsymbol{k}}^2,\\
&L_{\boldsymbol{k}, \boldsymbol{q}}^{( \pm)}=u_{\boldsymbol{k}+\boldsymbol{q}} u_{\boldsymbol{k}}^* \pm v_{\boldsymbol{k}+\boldsymbol{q}} v_{\boldsymbol{k}}^*, \quad M_{\boldsymbol{k}, \boldsymbol{q}}^{( \pm)}=u_{\boldsymbol{k}+\boldsymbol{q}}^* v_{\boldsymbol{k}}^* \pm v_{\boldsymbol{k}+\boldsymbol{q}}^* u_{\boldsymbol{k}}^*.
\end{aligned}    
\end{equation}
In the calculation, the laser pulse is modeled as 
\begin{equation}
\boldsymbol{A}(\boldsymbol{r}, t)=\boldsymbol{A}_0 \mathrm{e}^{-4 \ln (2)\left(\frac{t}{\tau}\right)^2} \cos \left(\Omega t-\boldsymbol{q}_0 \boldsymbol{r}\right) .    
\end{equation}
with the central frequency $\Omega$, wave momentum $\left|\boldsymbol{q}_0\right|=q_0=\frac{\Omega}{c}$ and full width at half maximum $\tau$ of the Gaussian envelope. \\
By writing down the equations of motion for the four expectation values, one gets a closed set of differential equations that can be numerically solved. The details of the implementation and solution of the equations of motion on a finite size grid in momentum space are specified in section Sec.~\ref{Sec:numerics_pp}.\\
\indent This formalism allows us to obtain the response of the superconductor to an external vector field that transfers a small momentum $q$. Since the transferred momentum of light is taken into account, the light mediates a coupling between fermions with off-diagonal momenta $({\boldsymbol{k}},\boldsymbol{k+q})$, such that this description goes beyond the Anderson pseudospin picture, where the coupling only happens between fermions with momenta $({\boldsymbol{k}},{\boldsymbol{k}})$. Thus, by adopting this formalism, we can couple to both the amplitude sector and the phase sector. \\
Furthermore, we can calculate the temporal evolution of the transient current density in the system as a function of the time delay between the pump and the probe pulse, see Appendix~\ref{app:eom}.\\
\indent Starting from the ground state of our system, we first couple the superconductor to the pump laser field with a finite momentum $\mathbf{q}_0$ and a fixed pulse duration $\tau_p $. This field is chosen to be much stronger than the probe field that will follow. Both pump and probe are Gaussian in time and do not overlap. 
 Within this setup, we varied the direction of the incident linearly polarized light field to study the dependence on the angle $\varphi$ of the vector field with respect to the ${\boldsymbol{k}}_x$-axis. Indeed, by varying the angle of the pulse, different oscillation symmetries can be triggered selectively~\cite{lukas}. 
By comparing the excitations resulting from the quantum quenches with respect to the one triggered by the pump pulse, we provide a mapping between the angle-resolved pump probe spectroscopy and the different oscillation symmetries of the condensate.

\section{Numerical Results}\label{Sec:numerics}
We present in this section the results of the numerical implementation of the quantum quenches and pump-probe described in Sec.~\ref{Sec:qq} and \ref{Sec:pp}  for the order parameter in Eq.~\eqref{eq:eta}. We show here the calculations for the d+id' case. The s+id and p+ip' cases are shown in Appendix~\ref{app:sid} and \ref{app:pip}.\\
The numerical spectra presented in the following sections are interpreted in light of the characteristics of the modes derived from the theoretical analysis.
\subsection{Quantum quench}\label{Sec:numerics_qq}
\begin{figure}[t!] 
\includegraphics[scale=.9]{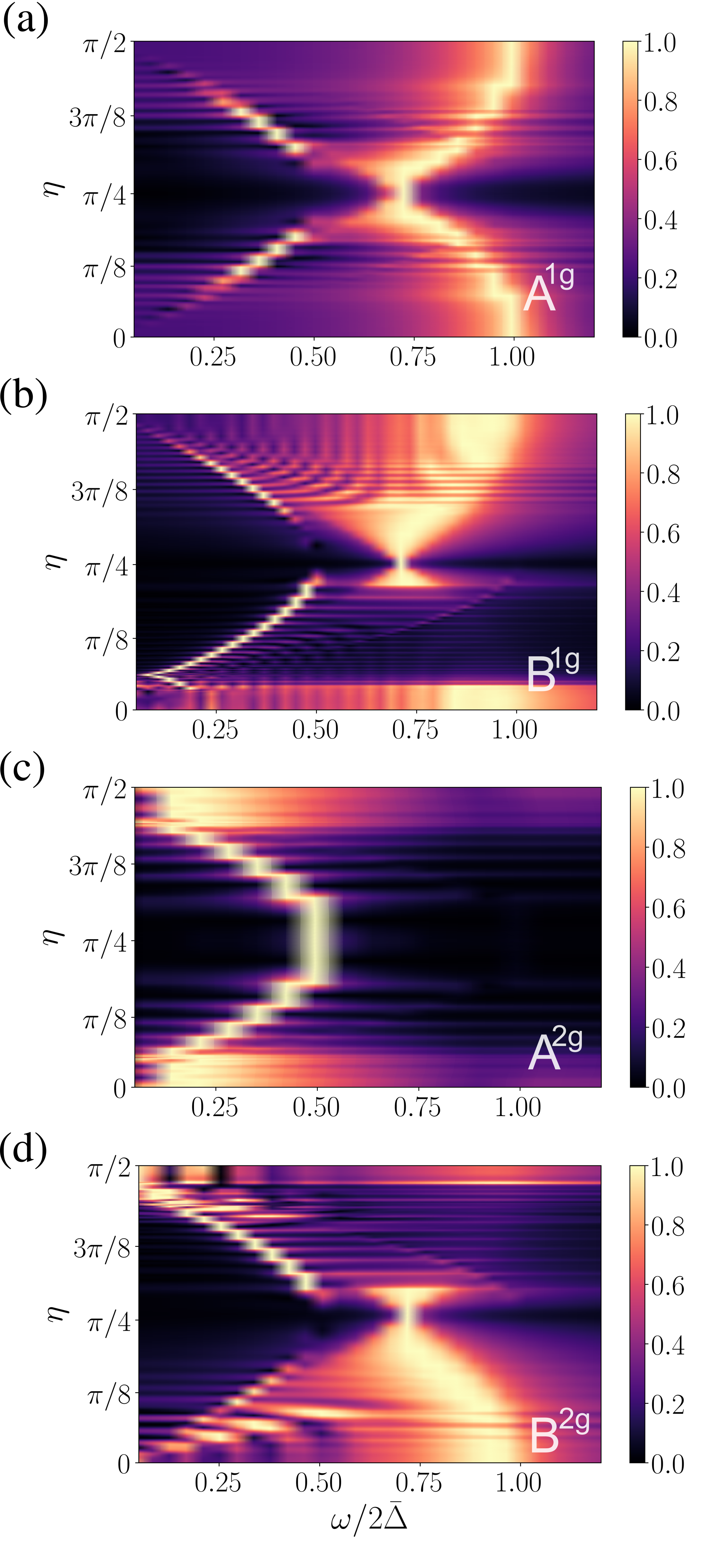} 
    \caption{Numerical simulation of the Higgs oscillations following a state quench with strength $\delta$=0.2 and Fourier spectrum $|\Delta(\omega)|=|FT|\Delta(t)||$ (in arbitrary units), as a function of frequency $\omega$ (scaled to the asymptotic value of the gap $2\bar{\Delta}$) and the mixing angle $\eta$, for a $d_{x^2 - y^2}+id_{xy}$ superconductor. From top to bottom: (a) $A^{1g}$-oscillations, (b) $A^{2g}$-oscillations, (c) $B^{1g}$-oscillations, (d) $B^{2g}$-oscillations.} 
    \label{fig:did_state_quench} 
\end{figure}

The Bloch equations for the Anderson pseudospins are solved by adopting a Runge-Kutta-4 algorithm on a two-dimensional grid in momentum space. We use a parabolic band dispersion, $\epsilon_{{\boldsymbol{k}}}=t_{hop}{{\boldsymbol{k}}}^2-\epsilon_F$, with parameters $t_{hop}= 50$~meV for the nearest-neighbor hopping and  Fermi energy $\epsilon_F$= 100~meV. The equilibrium gap value is chosen to be $|\Delta_{0}|$= 1~meV for all simulations. For the momentum grid, we only consider the region around the Fermi level within $\{ \bold{k}||\epsilon_{{\boldsymbol{k}}} | < \epsilon_c \}$, $\epsilon_c$ = 10~meV. Since all considered configurations for the quenches separate in radial and angular parts, we discretize the momentum space in polar coordinates: $\epsilon(k_x,k_y)\longrightarrow \epsilon(k,\phi)$, $\phi= \arctan(\frac{k_y}{k_x})$, such that in the radial direction we choose $N_{\boldsymbol{k}} = 2001$ points and in the angular direction $N_{\phi}=401$. The momentum resolution of the grid influences the maximum time up to which the calculation is possible, with the above choice of parameter $t_{max}\approx$ 50~psec. To perturb the system we then fix the strength of the quench to $\delta$ = 0.2 \cite{lukas,papenkort}.\\
\indent As described in Sec.~\ref{Sec:qq}, the quenches implemented here, according to Tab.~\ref{tab:quenches}, change the symmetry of the superconducting condensate with respect to the ground state symmetry, inducing oscillations that can be classified according to the irreducible representations of the point group of the lattice. We present in Fig.~\ref{fig:did_state_quench} the results obtained for the $d_{x^2-y^2}+id_{xy}$ case.\\ 
\indent The results display the Fourier transform of the oscillations plotted as a function of $\omega/2\bar{\Delta}$, where $\bar{\Delta}$ is the asymptotic value reached by the gap in the long time regime, which ultimately depends on the strength of the quench~\cite{Yuzbashyan_2006,Yuzbashyan_2005}. The two dimensional plots provide a picture of the dynamics obtained while changing the $\eta$ parameter introduced in Eq.~\eqref{eq:eta}, which represents the ratio between the two components.
In Fig.~\ref{fig:did_state_quench}, we observed the spectra of the d+id' state, we can appreciate the mirror symmetry with respect to $\eta = \pi/4$ for each plot, reflecting the symmetry of the condensate. In particular, for the $A^{1g}$ and $A^{2g}$ symmetry channels, we see that, as one would intuitively picture by looking at Fig.~\ref{fig:did_osc}, we mostly excite respectively the amplitude and the phase sector.  While in the latter, we can observe the dispersion of the relative phase mode, in the former, the two amplitude modes appear.  In particular, the relative amplitude and relative phase modes are sub-gap excitations of the system. In the $B^{1g}$ and $B^{2g}$ channels, the spectra are more involved, mirroring each other with respect to $\eta = \pi/4$. However, due to the complicated oscillation triggered, it is harder to disentangle the different branches. For the $B^{1g}$ case, for example, from 0 to $\eta=\pi/4$ the phase sector is mostly excited due to the type of deformation, which shifts the position of the nodes, induced on the $d_{x^2-y^2}$ component of the gap, which dominates the lower half of the plot. In the upper half of the plot, from $\eta=\pi/4$ to $\eta=\pi/2$, the $d_{xy}$ component becomes dominant and the amplitude sector is mostly excited; indeed, the quench mostly shifts weight inside the lobes. Moreover, other peaks appear in the spectra which are not ascribable to the normal modes of the system, as in Ref.~\cite{lukas}.  

\begin{figure}[t!] 
    \centering 
\includegraphics[scale=.6]{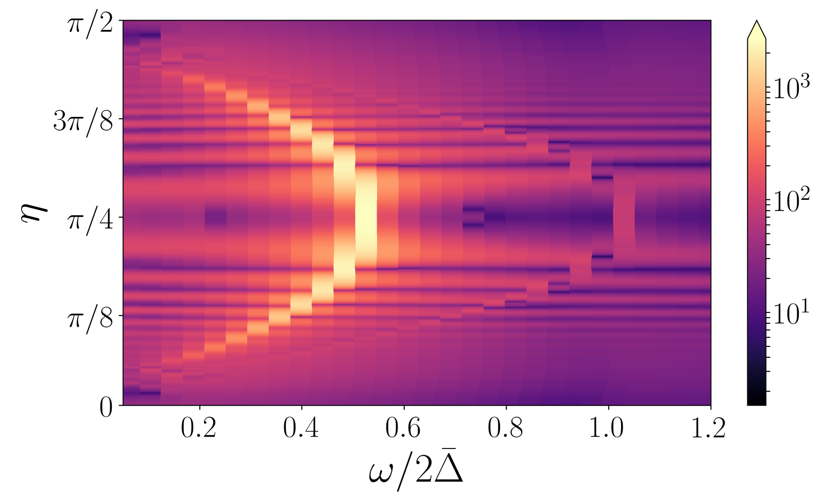} 
    \caption{ Logarithmic plot corresponding to d+id' oscillations in the $A^{2g}$-channel. In the 2D plots, in Fig.~\ref{fig:did_state_quench}(c), certain features are overborne by the normalization of the Fourier intensity. From the logarithmic scale, we can see the additional modes that are dynamically induced by the state quench as in Ref.\cite{lukas}.} 
    \label{fig:did_peaks} 
\end{figure}

Furthermore, to reveal the hidden features in the 2D plots presented above, we display in Fig.~\ref{fig:did_peaks} the same 2D plot as in Fig.~\ref{fig:did_state_quench}(c), but with a logarithmic scale. While in Fig.~\ref{fig:did_state_quench} the intensity of the Fourier transform is normalized for each line cut, here it is not. We can then observe that the intensity of the mode is strongest at $\eta=\pi/4$ and diminishes as it approaches the extreme values of $\eta=0,\pi$. Additionally, we see the extra mode dynamically induced by the asymmetric quench. 

\subsection{Pump-probe}\label{Sec:numerics_pp}
We now present the results obtained from the numerical implementation of the density matrix formalism described in Sec.~\ref{Sec:pp}, also in this case, the equations have been solved using a Runge-Kutta-4. The equations of motion are implemented on a different grid with respect to the Bloch equations in the previous section. Indeed, off-diagonal expectation values at \textbf{k} and $\mathbf{k^\prime}$ are coupled. This is due to the finite momentum $\mathbf{q}_0$ transferred from light which is coupling elements with momentum (\textbf{k},$\mathbf{k^\prime}$=\textbf{k}+n$\mathbf{q}_0$ ). As the coupling with more distant off-diagonal elements scales with the amplitude of the vector field $A_0$, terms higher than order n=4 for the pump-pulse (order n=1 for the probe-pulse) are neglected in the calculation \cite{papenkort,PhysRevB.90.014515}. To take the above into account, we choose a two-dimensional grid where the momentum space is discretized in the x-direction with a step size of $|\mathbf{q}_0|$, such to resolve the light momentum, and in y-axis with $N_y$ points \cite{PhysRevB.90.014515,lukas,papenkort}. The parameters used for the calculations are the following $\Delta_0=1$~meV, $t_{hop}=2000$~meV, $\epsilon_F = 10000$~meV, $\epsilon_c = 10$~meV, $N_y = 200$, $t_{max}\approx$ 50~psec. The parameters adopted for the vector potential are:
$\tau_p = 0.4$~ps, $|\mathbf{A}_p| = 7~10^{-8} \text{Js}~\text{C}^{-1} \text{m}^{-1} $ and $\hbar \omega = 2.2$~meV, the momentum $q_0$ transferred by light is then chosen as a quantization step $\delta k_x$ for the x-axis of the grid.
The equations are first solved by only considering the application of the pump-pulse. Then, we use a much weaker probe-pulse in the same direction to extract the transient optical conductivity in the system. The probe parameters are:
$\tau_p = 0.25$~ps, $|\mathbf{A}_p| = 1~10^{-8} \text{Js}~\text{C}^{-1} \text{m}^{-1} $ and $\hbar \omega = 1.85$~meV. The calculations are shown in the Appendix~\ref{app:oc}.
In this latter scenario, the equations of motion are integrated further on the same grid induced by the pump; indeed, for this calculation, we consider the momentum transferred by the probe negligible and we only couple off-diagonal elements up to nearest neighbors. We evolve the equations using the non-equilibrium state induced by the pump-pulse at a variable time $t$ as the initial state on which the probe-pulse is applied. Hereby, it is ensured that the pump and probe pulses do not overlap.\\

\begin{figure}[t!] 
    \centering 
 \includegraphics[scale=1]{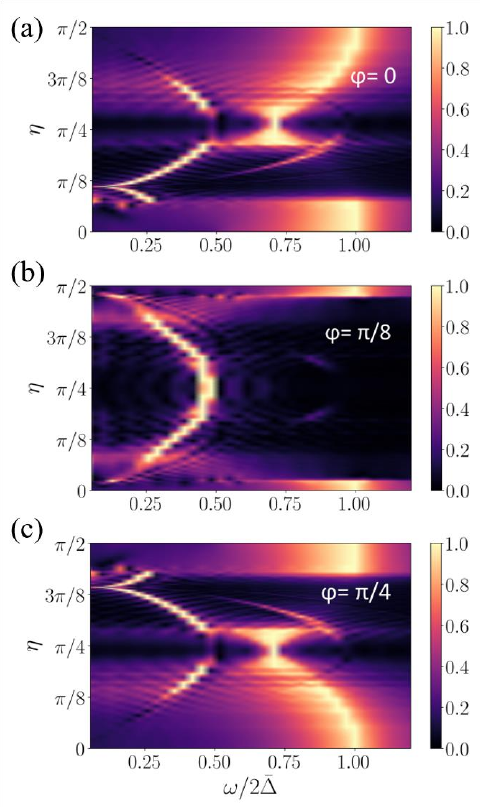} 
    \caption{Numerical simulation of the condensate oscillations for a d+id' order parameter following the application of a pump pulse at incident angles $\varphi=0, \varphi=\pi/8, \varphi=\pi/4$. Fourier spectrum $|\Delta(\omega)|=|FT|\Delta(t)||$ (in arbitrary units), as a function of frequency $\omega$ (scaled to the asymptotic value of the gap $2\bar{\Delta}$) and the mixing angle $\eta$.
    The pulse parameters are $\tau_p = 0.4$~ps, $|\mathbf{A}_p| = 7~10^{-8} \text{Js}~\text{C}^{-1} \text{m}^{-1} $ and $\hbar \omega = 4$~meV. For each $\eta$ value the spectrum of the oscillations is normalized to the maximum value in the visualized frequency range.} 
    \label{fig:did_pump} 
\end{figure}

\begin{figure}[htpb!] 
    \centering 
\includegraphics[scale=.6]{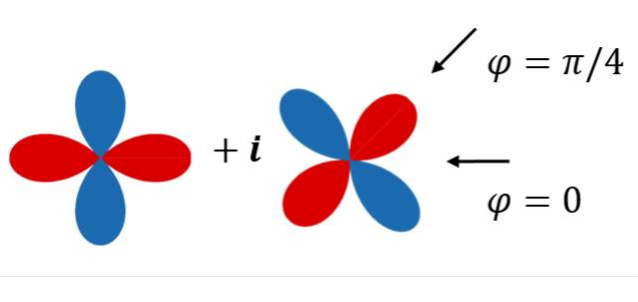} 
    \caption{ Schematic picture of a pump-pulse on d+id' condensate. At $\varphi=0$ the pump pulse hits the antinodal line of the first condensate while it is oriented to the nodal line of the second component. At $\varphi=\pi/4$ the picture is reversed.} 
    \label{fig:did_pumpprobe} 
\end{figure}

The results of the calculation are shown in Fig.~\ref{fig:did_pump}, and a schematic illustration of the pumping angle with respect to the condensate symmetry is shown in Fig.~\ref{fig:did_pumpprobe}.
In Fig.~\ref{fig:did_pump}, we illustrate the gap oscillations that occur following the pump pulse. Unlike the instantaneous quench described in the previous section, this pulse acts as a quench but lasts for a finite duration. To effectively function as a quench, the pulse duration is kept shorter than the intrinsic timescale of the dynamics it triggers. By varying the incident angle relative to the gap, we aim to induce asymmetric oscillations and map the resulting dynamics onto the group-theoretical classification discussed in Sec.~\ref{Sec:qq}.
As it appears from Fig.~\ref{fig:did_pump} the dynamics triggered by the pump-pulse depend on the pumping angle. By comparing with Fig.~\ref{fig:did_state_quench}, we build Tab.~\ref{tab:did_quenches}.
For example, in the case of $\varphi=0$, the pulse quenches the positive lobes of the $d_{x^2 -y^2}$ component differently compared to the negative lobes as the lobes' axes are parallel or perpendicular to the pulse directions. For the same incident angle, however, the effect on the second component, the $d_{xy}$, is different, as the pulse is aligned in between both lobes, along the nodal lines. The dynamics triggered is reminiscent of the one observed for oscillations in the $B^{1g}$, see Fig.~\ref{fig:did_state_quench}(b). Analogous reasoning holds for the other angles, $\varphi=\pi/8,\pi/4$.
{\def\arraystretch{2}\tabcolsep=3 pt
\begin{table}[htpb!] 
    \centering
    \begin{tabular}{c|c|c}
    \toprule
        Gap & Condensate osc& Pump angle  \\\midrule
        
        $d_{x^2 - y^2} + id_{xy}$ &  $A^{1g}_{d+id'}$ & -\\
                &  $B^{1g}_{d+id'}+A^{1g}_{d+id'}$ & 0\\
                &  $A^{2g}_{d+id'}+A^{1g}_{d+id'}$ & $\pi/8$\\
                & $B^{2g}_{d+id'}+A^{1g}_{d+id'}$ & $\pi/4$\\
                \bottomrule
                
    \end{tabular}
    \caption{
    Comparison between pump-probe and group theory analysis for the d+id' case. The second column lists all fundamental condensate oscillations. The third column lists the pumping angle able to induce the corresponding oscillation.}
    \label{tab:did_quenches} 
\end{table}
}\\
\indent We note here that in Fig.~\ref{fig:did_pump} the spectra at $\varphi=0$ and $\varphi=\pi/4$ present, in the low frequency region, some anomalous behavior around $\pi/8$ and $3\pi/8$ respectively. This is due to the strength of the pump and to the size of the grid; indeed, along the above-mentioned line cuts, the strength of the pump is such that the perturbation induced is comparable with the size of the gap minima. It was our choice, indeed, to maintain the strength of the perturbation constant throughout the entire calculation. This artifact, ultimately, depends on the numerical implementation. The behavior is indeed different in the case of the p+ip'-wave shown in the Appendix~\ref{app:pip}.

\section{Conclusion}

To summarize, we have employed a variety of techniques to study the collective modes in two-dimensional time-reversal symmetry-breaking superconductors, specifically in the cases of s+id, d+id', and p+ip'. We first analyzed the bosonic spectrum within the Landau-Ginzburg framework, obtaining insightful information on the dispersion of the excitations present in the systems in terms of the relative magnitude of the two components of the gap. We then studied the stability of the modes by performing a microscopic calculation in the clean limit. Within this analysis, we obtained information on the way these excitations couple to a vector field. In particular, while the amplitude modes couple to the light field as the Higgs mode, that is,  $\left(\partial^2_{\boldsymbol{k}} \epsilon_{{\boldsymbol{k}}}\right) \epsilon_{{\boldsymbol{k}}}$, the relative phase fluctuation couple as the Leggett/Bardasis-Schrieffer mode, that is, $\partial^2_{\boldsymbol{k}} \epsilon_{{\boldsymbol{k}}}$~. In general, the relative importance of the modes in the spectrum of a realistic system is going to depend on the band structure.

We introduced a classification scheme for non-equilibrium oscillations in the cases of s+id, d+id', and p+ip', which allows us to characterize the ground state of superconducting condensates. Our calculations show that, depending on the symmetry of the quench and the gap function, we can trigger different oscillations in the condensate that then lead to a specific response of the excitations. It then follows that fluctuations in the relative amplitude and phase, which are sub-gap excitations, can be potentially selectively triggered by inducing different oscillations on the condensate.
The link between the pump-probe framework and the group theoretical classification is explicated in Fig.~\ref{fig:conclusion}. The cartoon summarizes the core results of the numerical analysis in the paper, illustrating the idea that by changing the direction of the pump pulse, we can trigger the oscillations of the superconducting condensate in a specific symmetry channel and hence enhance the response of certain excitations. In particular, for the case of the $d_{x^2-y^2}+i d_{xy}$ shown in the main text, we see that the application of a pulse at $\varphi=\pi/8$ results in the enhancement of the response in the phase sector, while in other configurations both the amplitude and phase sectors are triggered giving rise to more complicated spectra. For the scenario, modeled by the linear pseudospin in the main text, where the coupling with the light transferred momentum is negligible, the oscillation triggered is in the $A^{1g}$ channel and the response is dominated by the amplitude sector.  

Measurement of these bosonic subgap excitations could provide key evidence of the T-broken state, in line with the new field of collective mode spectroscopy. Our work investigates the behavior of the excitations in these systems in an out-of-equilibrium scenario, paving the way for the experimental detection of these “generalized clapping modes” within ultrafast and non-linear optical experiments, extending the concept of collective mode spectroscopy to another class of unconventional superconductors. The classification and characterization of the oscillations depend both on the orbital symmetries of the two order parameter components and the relative amplitude of the two components, opening the possibility of inferring essential features of the ground state symmetry by performing spectroscopic studies with momentum-dependent probes~\cite{PhysRevLett.109.147403,Gedik2003}.

It is important to note that we performed the calculations in a simplified setting, without incorporating additional energy scales or other degrees of freedom, such as subdominant channels or competing orders. In real bulk materials where time-reversal symmetry-breaking scenarios may be favorable, the system is typically more complex, often exhibiting a multiband structure and strong electronic correlations. These features, along with the presence of disorder, significantly influence and modify the collective excitation spectra. Understanding these effects is crucial for accurately describing the physical properties and emergent phenomena in such superconductors. 

\section*{ACKNOWLEDGMENTS}
We thank Haruki Matsumoto for his contributions to Sec.~\ref{Sec:linpseudo}. We thankfully acknowledge discussions with Sida Tian and Jakob Dolgner.

\appendix
\section{Linear analysis with pseudospin model}\label{app:linearpseudo}
We here make use of the simplicity of the BCS Hamiltonian in the linearized pseudospin representation for $T=0$. 
For the tetragonal system in case of an OP of the form $\Delta_{\boldsymbol{k}}=\Delta_{1,\boldsymbol{k}} + i\Delta_{2,\boldsymbol{k}}$, the linearized equations of motion in the frequency domain are
\begin{equation}
\resizebox{\columnwidth}{!}{$
 \begin{aligned}
    i\omega \delta\sigma^{x}_{{\boldsymbol{k}}}&= \frac{\epsilon_{{\boldsymbol{k}}}}{E_{{{\boldsymbol{k}}},eq}}\delta\Delta^{''}_{{\boldsymbol{k}}}  
    - 2\Delta^{''}_{{{\boldsymbol{k}}},eq}\delta\sigma^{z}_{{\boldsymbol{k}}} 
    - \frac{\Delta^{''}_{{{\boldsymbol{k}}},eq}}{E_{{{\boldsymbol{k}}},eq}}\delta b^{z}_{{\boldsymbol{k}}} 
    - 2 \epsilon_{{\boldsymbol{k}}} \delta \sigma^{y}_{{\boldsymbol{k}}} , \\
     i\omega \delta\sigma^{y}_{{\boldsymbol{k}}}&= \frac{\Delta^{'}_{{{\boldsymbol{k}}},eq}}{E_{{{\boldsymbol{k}}},eq}}\delta b^{z}_{{\boldsymbol{k}}} 
     + 2 \epsilon_{{\boldsymbol{k}}} \delta \sigma^{x}_{{\boldsymbol{k}}} 
     - \frac{\epsilon_{{\boldsymbol{k}}}}{E_{{{\boldsymbol{k}}},eq}}\delta\Delta^{'}_{{\boldsymbol{k}}} 
     + 2\Delta^{'}_{{{\boldsymbol{k}}},eq} \delta \sigma^{z}_{{\boldsymbol{k}}}, \\
     i\omega \delta\sigma^{z}_{{\boldsymbol{k}}}&= 
     -\frac{\Delta^{''}_{{{\boldsymbol{k}}},eq}}{E_{{{\boldsymbol{k}}},eq}}\delta \Delta^{'}_{{\boldsymbol{k}}}
     - 2\Delta^{'}_{{{\boldsymbol{k}}},eq}\delta\sigma^{y}_{{\boldsymbol{k}}} 
     + \frac{\Delta^{'}_{{{\boldsymbol{k}}},eq}}{E_{{{\boldsymbol{k}}},eq}}\delta \Delta^{''}_{{\boldsymbol{k}}}
     + 2\Delta^{''}_{{{\boldsymbol{k}}},eq}\delta\sigma^{x}_{{\boldsymbol{k}}}, 
\end{aligned}$}
\tag{A1}
\end{equation}

with 
\begin{equation}
\begin{aligned}
&\delta\boldsymbol{\sigma_{k}}(\omega) = \boldsymbol{\sigma_{k}} - \boldsymbol{\sigma}^{eq}_{{\boldsymbol{k}}},\\
&\delta\Delta^{'}_{\boldsymbol{k}} = \delta\Delta^{'}_{1,\boldsymbol{k}}(\omega) + \delta\Delta^{'}_{2,\boldsymbol{k}}, \quad
\delta\Delta^{''}_{\boldsymbol{k}} = \delta\Delta^{''}_{1,\boldsymbol{k}}(\omega) + \delta\Delta^{''}_{2,\boldsymbol{k}},\\
&\delta\Delta^{'}_{1,\boldsymbol{k}} = \Delta^{'}_{1,\boldsymbol{k}}(\omega) - \Delta^{eq}_{1,\boldsymbol{k}}, \quad
\delta\Delta^{''}_{1,\boldsymbol{k}} = \Delta^{''}_{1,\boldsymbol{k}}(\omega) - 0,\\
&\delta\Delta^{'}_{2,\boldsymbol{k}} = \Delta^{'}_{2,\boldsymbol{k}}(\omega) - 0, \quad
\delta\Delta^{''}_{2,\boldsymbol{k}} = \Delta^{''}_{2,\boldsymbol{k}}(\omega) - \Delta^{eq}_{2,\boldsymbol{k}}.
\end{aligned}
\tag{A2}
\end{equation}

The effect of the laser is given by:
\begin{equation}
\delta b_{{\boldsymbol{k}}}^z(t) = b_{{\boldsymbol{k}}}^z(t) - \epsilon_{{\boldsymbol{k}}} 
\simeq \left(\frac{e^2}{2} \right) \sum_{i j} 
\left(\partial_{k_i} \partial_{k_j} \epsilon_{{\boldsymbol{k}}}\right) A_i(t) A_j(t).    
\tag{A3}
\end{equation}

By solving self-consistently linearized equations, we obtain the following equations for the fluctuations:
\begin{equation}
\left(\begin{array}{c}
\delta\Delta^{'}_{{\boldsymbol{k}},1} \\
\delta\Delta^{"}_{{\boldsymbol{k}},1}\\
\delta\Delta^{'}_{{\boldsymbol{k}},2}\\
\delta\Delta^{"}_{{\boldsymbol{k}},2}
\end{array}\right) 
= M 
\left(\begin{array}{c}
\delta\Delta^{\prime}_{{\boldsymbol{k}},1} \\
\delta\Delta^{\prime\prime}_{{\boldsymbol{k}},1}\\
\delta\Delta^{\prime}_{{\boldsymbol{k}},2}\\
\delta\Delta^{\prime\prime}_{{\boldsymbol{k}},2}
\end{array}\right) 
+ N 
\left(\begin{array}{c}
\Delta^{eq,\prime}_{{\boldsymbol{k}},1} \\
\Delta^{eq,\prime\prime}_{{\boldsymbol{k}},1}\\
\Delta^{eq,\prime}_{{\boldsymbol{k}},2}\\
\Delta^{eq,\prime\prime}_{{\boldsymbol{k}},2}
\end{array}\right)
\tag{A4}
\end{equation}
with\\

\begin{multline*} 
 \noindent \text{M}=\\
 \resizebox{\columnwidth}{!}{$
\left[
\begin{matrix} -2(I_{11} + F_{1111} \Delta^{'}_{1}{}^2 + F_{1122} \Delta^{'}_{2}{}^2 ) V_1 & 2 F_{1122} (\Delta^{''}_{2} \Delta^{'}_{1} + \Delta^{''}_{1} \Delta^{'}_{2}) V_1 \\ 2 F_{1122} (\Delta^{''}_{2} \Delta^{'}_{1} + \Delta^{''}_{1} \Delta^{'}_{2}) V_2 & -2(I_{22} + F_{2222} \Delta^{''}_{2}{}^2 + F_{1122} \Delta^{''}_{1}{}^2 ) V_2 \\ 2 (F_{1111}\Delta^{''}_{1} \Delta^{'}_{1} + F_{1122}\Delta^{''}_{2} \Delta^{'}_{2}) V_1 & -4 F_{1122}\Delta^{'}_{2} \Delta^{'}_{1} V_1\\ -4 F_{1122}\Delta^{''}_{2} \Delta^{''}_{1} V_2 & 2 (F_{1122}\Delta^{''}_{1} \Delta^{'}_{1} + F_{2222}\Delta^{''}_{2} \Delta^{'}_{2}) V_2 \end{matrix}
\right.
$}
\\
 \resizebox{\columnwidth}{!}{$
\left.
\begin{matrix} 2 (F_{1111}\Delta^{''}_{1} \Delta^{'}_{1} + F_{1122}\Delta^{''}_{2} \Delta^{'}_{2}) V_1 & -4 F_{1122}\Delta^{'}_{2} \Delta^{'}_{1} V_1 \\ -4 F_{1122}\Delta^{''}_{2} \Delta^{''}_{1}V_2 & 2 (F_{1122}\Delta^{''}_{1} \Delta^{'}_{1} + F_{2222}\Delta^{''}_{2} \Delta^{'}_{2}) V_2 \\ -2(I_{11} + F_{1111} \Delta^{''}_{1}{}^2 + F_{1122} \Delta^{''}_{2}{}^2 ) V_1 & 2 F_{1122} (\Delta^{''}_{2} \Delta^{'}_{1} + \Delta^{''}_{1} \Delta^{'}_{2}) V_1 \\ 2 F_{1122} (\Delta^{''}_{2} \Delta^{'}_{1} + \Delta^{''}_{1} \Delta^{'}_{2}) V_1 & -2(I_{22} + F_{2222} \Delta^{'}_{2}{}^2 + F_{1122} \Delta^{'}_{1}{}^2 ) V_1 \end{matrix}
\right]
$}
\end{multline*}
where we dropped the superscript "eq" from the $\Delta_1,\Delta_2$,
and\\[0.5em]
\indent\text{N}=\\
[1em]
\resizebox{\columnwidth}{!}{$
\begin{bmatrix}
     A^2 e^2 \mathrm{X}_{11} V_1 & 0 & -\frac{1}{2} i A^2 e^2 \mathrm{Y}_{11} V_1 \omega & 0 \\ 0 & A^2 \mathrm{e}^2 \mathrm{X}_{22} V_{2} & 0 & \frac{1}{2} i A^2 \mathrm{e}^2 \mathrm{Y}_{22} V_2 \omega \\ \frac{1}{2} i A^2 \mathrm{e}^2 \mathrm{Y}_{11} V_1 \omega & 0 & \mathrm{~A}^2 \mathrm{e}^2 \mathrm{X}_{11} V_1 & 0 \\ 0 & -\frac{1}{2} \text { i } A^2 \mathrm{e}^2 \mathrm{Y}_{22} V_2 \omega & 0 & \mathrm{~A}^2 \mathrm{e}^2 \mathrm{X}_{22} V_2 
\end{bmatrix}$}\\[1em]
where\\
\begin{equation}
    \begin{aligned}
       &F_{ijlm}=\frac{1}{D} \sum_{{\boldsymbol{k}}} \frac{  f^{i}_{{\boldsymbol{k}}} f^{j}_{{\boldsymbol{k}}} f^{l}_{{\boldsymbol{k}}} f^{m}_{{\boldsymbol{k}}}}{E_{{\boldsymbol{k}}} \left(4 E^{2}_{{\boldsymbol{k}}}-\omega^2\right)} ,\\
       &I_{ij}= \frac{1}{D} \sum_{{\boldsymbol{k}}} \frac{ f_{{\boldsymbol{k}}}^i f_{{\boldsymbol{k}}}^j}{E_{{\boldsymbol{k}}} \left(4 E^{2}_{{\boldsymbol{k}}}-\omega^2\right)}\epsilon_{{\boldsymbol{k}}}^2,\\
       &Y_{ij}= \frac{1}{D} \sum_{{\boldsymbol{k}}}  \frac{ f_{{\boldsymbol{k}}}^i f_{{\boldsymbol{k}}}^j}{E_{{\boldsymbol{k}}} \left(4 E^{2}_{{\boldsymbol{k}}}-\omega^2\right)}\frac{\partial^2 \epsilon_{{\boldsymbol{k}}}}{\partial k^{2}_x},\\
       &X_{ij}= \frac{1}{D} \sum_{{\boldsymbol{k}}}  \frac{f_{{\boldsymbol{k}}}^i f_{{\boldsymbol{k}}}^j}{E_{{\boldsymbol{k}}} \left(4 E^{2}_{{\boldsymbol{k}}}-\omega^2\right)}\frac{\partial^2 \epsilon_{{\boldsymbol{k}}}}{\partial k^{2}_x} \epsilon_{{\boldsymbol{k}}},
    \end{aligned}
\tag{A5}
\end{equation}
where D is the density of states assumed to be constant around the Fermi surface.
In particular, $X_{ij}$ and $Y_{ij}$ contain the dependence on the band dispersion. We can expand the above equations to write in a compact form the expressions for the gap components:

\begin{equation}
\begin{aligned}
        &\delta \Delta^{'}_i=-V_i \int_{-\epsilon_c}^{\epsilon_{\mathrm{c}}} \mathrm{d} \epsilon \int_{0}^{2\pi} \mathrm{d}\varphi f^i_{{\boldsymbol{k}}} \frac{1}{E_{{\boldsymbol{k}}} (4 E^2_{{{\boldsymbol{k}}}}-\omega^2)} \\
        &\{(f^{1}_{{\boldsymbol{k}}} \delta\Delta^{'}_1+f^{2}_{{\boldsymbol{k}}} \delta\Delta^{'}_2)(+2 f^{2}_{{\boldsymbol{k}}} f^{2}_{{\boldsymbol{k}}}\Delta^{''}_{2,eq}{}^2+2\epsilon^{2}_{{\boldsymbol{k}}})\\
        &+(f^{1}_{{\boldsymbol{k}}} \delta\Delta^{''}_1+f^{2}_{{\boldsymbol{k}}} \delta\Delta^{''}_2)(-2 f^{1}_{{\boldsymbol{k}}} f^{2}_{{\boldsymbol{k}}}\Delta^{'}_{1,eq}\Delta^{''}_{2,eq})\\+&\frac{e^2 A^2}{2} \frac{\partial^2 \epsilon_{{\boldsymbol{k}}}}{\partial k^{2}_x} (-2 f^{1}_{{\boldsymbol{k}}}\Delta^{'}_{1,eq}\epsilon_{{\boldsymbol{k}}}-i \omega f^{2}_{{\boldsymbol{k}}}\Delta^{''}_{2,eq})\}
       ,\\[1.5ex]
        &\delta \Delta^{''}_i=-V_i \int_{-\epsilon_c}^{\epsilon_{\mathrm{c}}} \mathrm{d} \epsilon \int_{0}^{2\pi} \mathrm{d}\varphi f^i_{{\boldsymbol{k}}} \frac{1}{ E_{{\boldsymbol{k}}} (4 E^2_{{{\boldsymbol{k}}}}-\omega^2)} \\
        &\{(f^{1}_{{\boldsymbol{k}}} \delta\Delta^{'}_1+f^{2}_{{\boldsymbol{k}}} \delta\Delta^{'}_2)(-2 f^{2}_{{\boldsymbol{k}}} f^{1}_{{\boldsymbol{k}}}\Delta^{'}_{1,eq}\Delta^{''}_{2,eq})\\
        &+(f^{1}_{{\boldsymbol{k}}} \delta\Delta^{''}_1+f^{2}_{{\boldsymbol{k}}} \delta\Delta^{''}_2)(2 f^{1}_{{\boldsymbol{k}}} f^{1}_{{\boldsymbol{k}}} \Delta^{'}_{1,eq}{}^2+2\epsilon^{2}_{{\boldsymbol{k}}})\\
        &+\frac{e^2 A^2}{2} \frac{\partial^2 \epsilon_{{\boldsymbol{k}}}}{\partial k^{2}_x} (-2 f^{2}_{{\boldsymbol{k}}}\Delta^{''}_{2,eq}\epsilon_{{\boldsymbol{k}}} + i \omega f^{1}_{{\boldsymbol{k}}}\Delta^{'}_{1,eq})\}.
\end{aligned}\label{eq:fluct}
\tag{A6}
\end{equation}

By looking at the equations above we can notice that while the amplitude fluctuations couple to the vector potential via $\left(e^2 A^2 / 2\right) \left(\partial^2_{k_x} \epsilon_{{\boldsymbol{k}}}\right) \epsilon_{{\boldsymbol{k}}} $, as the Higgs mode, such that the response is expected to be absent in the clean and parabolic band case; the phase fluctuations couple via $\left(e^2 A^2 / 2\right) \left(\partial^2_{k_x} \epsilon_{{\boldsymbol{k}}}\right)$, e.g. as the case for the Leggett mode, the response is then expected to be strong.\\

Keeping in mind that $f^{1}_{{\boldsymbol{k}}}$ and $f^{2}_{{\boldsymbol{k}}}$ are orthogonal, it is possible to simplify the above equations and,
by inverting them, one obtains the final equations in the form of 
 \begin{equation}
\left(\begin{array}{c}
\delta\Delta^{'}_{1}(\omega) \\
\delta\Delta^{"}_{1}(\omega)\\
\delta\Delta^{'}_{2}(\omega)\\
\delta\Delta^{"}_{2}(\omega)
\end{array}\right) =\frac{e^2 A^2(\omega)}{2} \left(\begin{array}{c}
H_{1}(\omega) \\
h_{1}(\omega)\\
H_{2}(\omega)\\
h_{2}(\omega)
\end{array}\right)
\tag{A7}
 \end{equation}
the functions $H_{\gamma=1,2}(\omega), h_{\gamma=1,2}(\omega)$ contains the resonances of the collective modes in the system, the full response of the gap to the applied electric field is then given by $|(H_1(\omega)+H_2(\omega)) + i (h_1(\omega)+h_2(\omega))|$. 

\section{Coupling to vector potential and equations of motion}\label{app:eom}
Within the density matrix formalism, the light pulse is described by a vector potential $\mathbf{A}(\mathbf{r},t)$.
 Here, we work in a gauge in which the scalar potential $\phi(\mathbf{r}, t)$ is set to zero. The vector potential enters the Hamiltonian via minimal coupling $\mathbf{p} \rightarrow \mathbf{p} + e\mathbf{A}(\mathbf{r}, t)$ with the electron charge e. Assuming a quadratic dispersion, the BCS Hamiltonian in second quantization coupled to the vector potential reads

\begin{align}
H &=H_{\mathrm{BCS}}+H_{\mathrm{em}}^{(1)}+H_{\mathrm{em}}^{(2)},\tag{B1} \\
H_{\mathrm{em}}^{(1)} & =\frac{e \hbar}{2 m} \sum_{{\boldsymbol{k}}, \boldsymbol{q}, \sigma}(2 {\boldsymbol{k}}+\boldsymbol{q}) \cdot \boldsymbol{A}_{\boldsymbol{q}}(t) c_{{\boldsymbol{k}}+\boldsymbol{q}, \sigma}^{\dagger} c_{{\boldsymbol{k}}, \sigma},\tag{B2}  \\
H_{\mathrm{em}}^{(2)} & =\frac{e^2}{2 m} \sum_{{\boldsymbol{k}}, \boldsymbol{q}, \sigma}\left[\sum_{\boldsymbol{q}^{\prime}} \boldsymbol{A}_{\boldsymbol{q}-\boldsymbol{q}^{\prime}}(t) \cdot \boldsymbol{A}_{\boldsymbol{q}^{\prime}}(t)\right] c_{{\boldsymbol{k}}+\boldsymbol{q}, \sigma}^{\dagger} c_{{\boldsymbol{k}}, \sigma}.\tag{B3} 
\end{align}

Since we are concerned with the effects of a short and intense THz laser pulse, we retain up to the second order term in the interaction Hamiltonian. To perform the calculation we then perform a Bogoliubov transformation on $H$ and we derive the equations for the Bogoliubov expectation values in Sec.~\ref{Sec:pp}, see \cite{papenkort}.

\subsection{Optical conductivity}\label{app:oc}

To calculate the optical conductivity, we calculate the temporal evolution of the current density as a function of the time delay between the quench and probe pulse.
In linear response, the optical conductivity $\sigma(\omega)$ is defined by
\begin{align}
\mathbf{j}_{\boldsymbol{q}}\left(\omega, t_{\mathrm{pp}}\right) = 
-\mathrm{i} \omega \sigma_{\boldsymbol{q}}\left(\omega, t_{\mathrm{pp}}\right) \boldsymbol{A}_{\boldsymbol{q}}\left(\omega, t_{\mathrm{pp}}\right) 
\tag{B4}
\end{align}

with $\boldsymbol{A}_{\boldsymbol{q}}(\omega)=A_{\boldsymbol{q}}(\omega) \hat{e}_A$ and $\left|\hat{e}_A\right|=1$, we obtain for the optical conductivity
\begin{align}
\sigma_{\boldsymbol{q}}\left(\omega, t_{\mathrm{pp}}\right) = 
\frac{\mathbf{j}_{\boldsymbol{q}}\left(\omega, t_{\mathrm{pp}}\right) \hat{e}_A}
{-\mathrm{i} \omega A_{\boldsymbol{q}}\left(\omega, t_{\mathrm{pp}}\right)}
\tag{B5}
\end{align}

In this expression, the pump-probe delay $t_{\mathrm{pp}}$ is a variable parameter, for which the calculation is repeated. This time delay is chosen such that pump and probe pulse do not overlap.

The current $\mathbf{j}_{\boldsymbol{q}}(t)=\mathbf{j}^{(1)}_{\boldsymbol{q}}(t)+\mathbf{j}^{(2)}_{\boldsymbol{q}}(t)$ consists of two terms, where the second term $\mathbf{j}^{(2)}_{\boldsymbol{q}}(t) \propto A(t)$ can be neglected as it only leads to an offset in the imaginary part of the conductivity \cite{papenkort,lukas}.
The first term expressed in the Bogoliubov quasiparticles reads
\begin{align}
\mathbf{j}^{(1)}_{\boldsymbol{q}} = &-\frac{e \hbar}{2 m V} \sum_{{\boldsymbol{k}}} 
(2 {\boldsymbol{k}}+\boldsymbol{q}) 
\Big( L^{(+) *} \alpha_{{\boldsymbol{k}}}^{\dagger} \alpha_{{\boldsymbol{k}}+\boldsymbol{q}} 
- L^{(+)} \beta_{{\boldsymbol{k}}+\boldsymbol{q}}^{\dagger} \beta_{{\boldsymbol{k}}} \notag \\
&\quad - M^{(-) *} \alpha_{{\boldsymbol{k}}}^{\dagger} \beta_{{\boldsymbol{k}}+\boldsymbol{q}}^{\dagger} 
- M^{(-)} \alpha_{{\boldsymbol{k}}+\boldsymbol{q}} \beta_{{\boldsymbol{k}}} \Big) .
\tag{B6}
\end{align}

\begin{figure}[htpb!] 
    \centering 
 \includegraphics[scale=0.4]{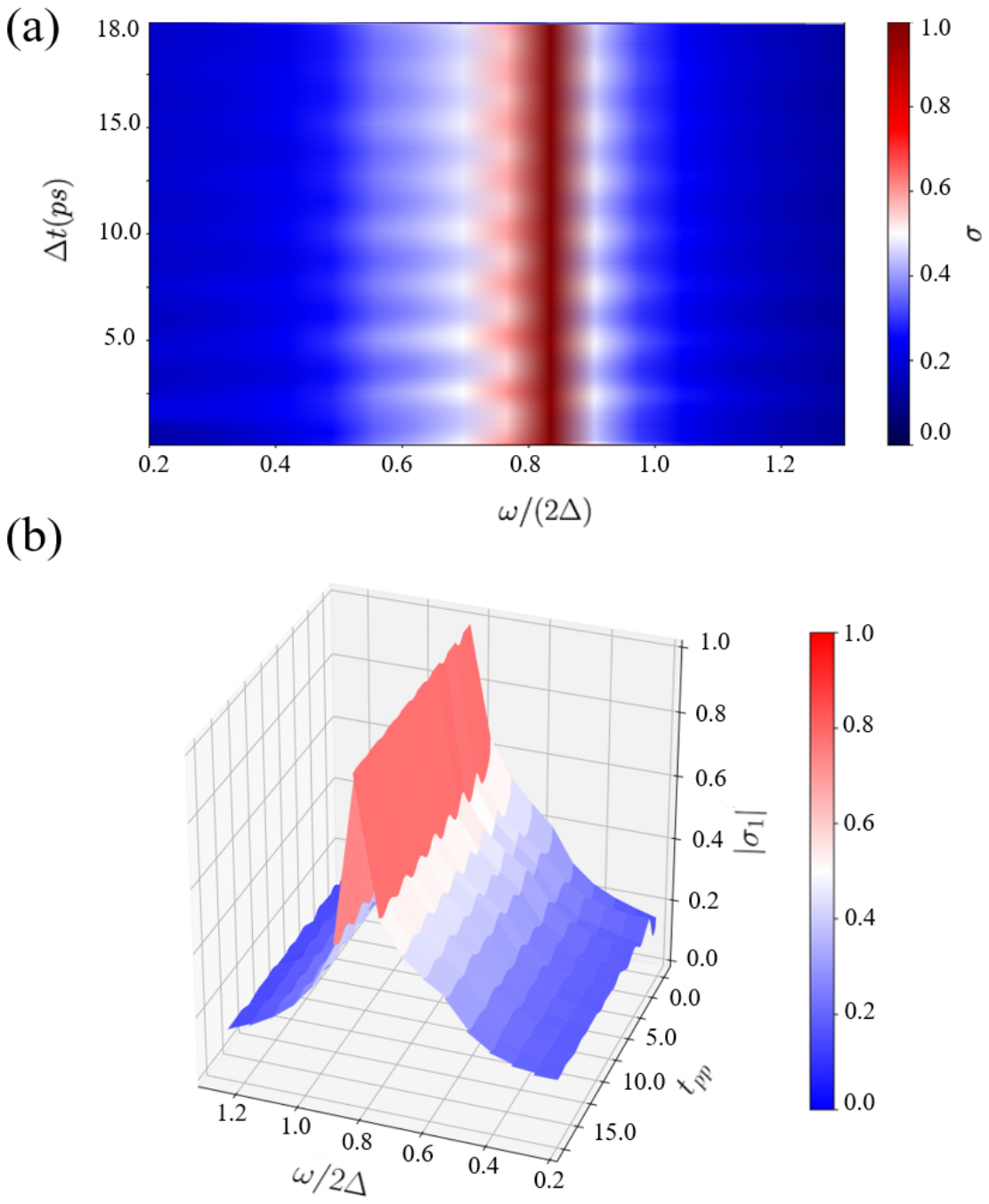} 
    \caption{Plot of $| Re~ \sigma(\Delta t_{pp}, \omega)|$  for a d+id' with $\eta=\pi/3$ irradiated at an angle of $\varphi=0$ after the application of a probe pulse with parameters: $\tau_p = 0.4$~ps, $|\mathbf{A}_p| = 7~10^{-8} \text{Js} \text{C}^{-1} \text{m}^{-1} $ and $\hbar \omega = 4$~meV for a d+id' superconductor. The spectrum is normalized to the maximum value in the visualized frequency range. It is possible to identify two peaks corresponding to the two amplitude modes.} 
    \label{fig:current} 
\end{figure}

From the same numerical simulation in Sec.~\ref{Sec:numerics_pp} we extract the current. In Fig.~\ref{fig:current} we show the real part of the transient optical conductivity for a given angle $\varphi$ and a fixed delay time $t_{pp}$, as a function of $\eta$ and $\omega$ for the d+id' case. The signal oscillates with respect to the time delay between the quench and probe pulse, reflecting the excitation of the modes. 

\section{s+id -wave superconductor}\label{app:sid}
\begin{figure}[htpb!] 
    \centering 
\includegraphics[scale=.8]{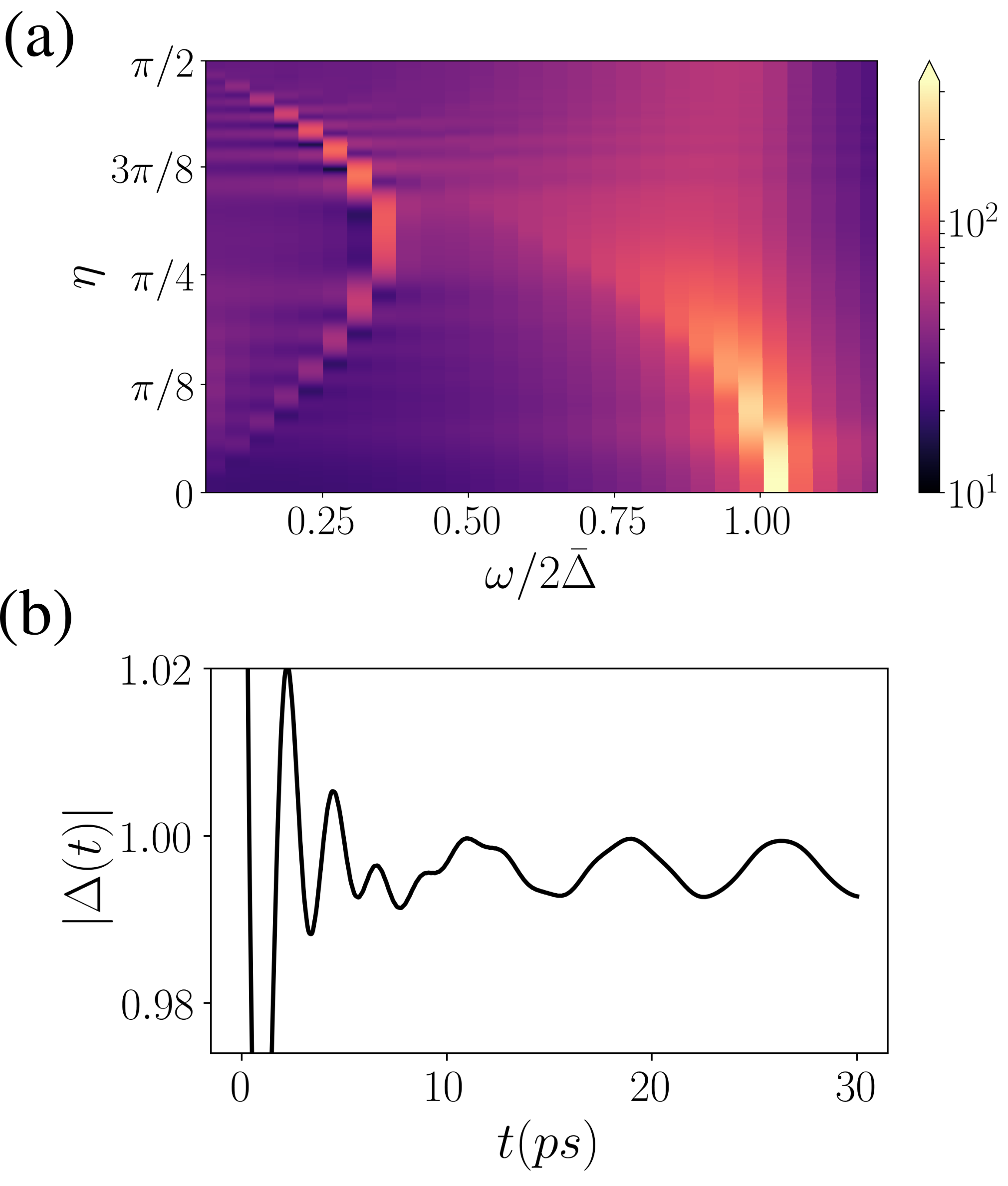} 
    \caption{ Numerical simulation of the oscillations following a state quench with strength $\delta$=0.2 and Fourier spectrum $|\Delta(\omega)|=|FT|\Delta(t)||$ (in arbitrary units), as a function of frequency $\omega$ (scaled to the asymptotic value of the gap $2\bar{\Delta}$) and the mixing angle $\eta$ for a $s+id_{x^2 - y^2}$ superconductor in the $A^{1g}$-channel. (a) Collective spectra of the oscillation (b) oscillation in time of the gap corresponding to a line cut of the plot in (a) at $\eta=\pi/3$.}  
    \label{fig:sup_sid} 
\end{figure}

\begin{figure}[t!] 
    \centering 
 \includegraphics[scale=0.325]{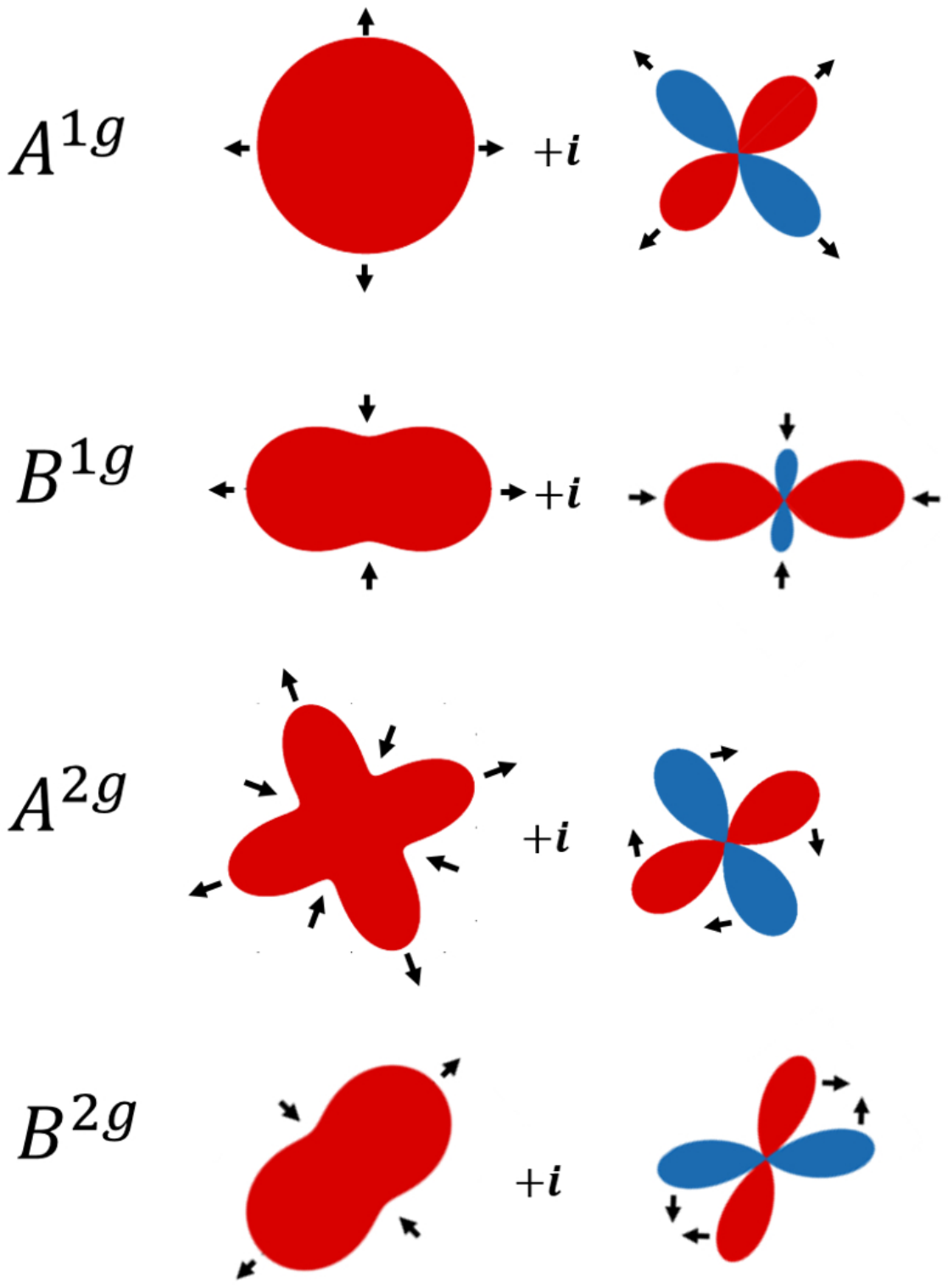} 
    \caption{Possible even oscillation symmetries for $s+id_{x^2 - y^2} $ superconductor with point group symmetry $D_{4h}$ of the underlying lattice. The arrows represent the motion of the lobes over time.} 
 ,   \label{fig:sid_osc}
\end{figure}

\begin{figure}[htpb!] 
    \centering 
\includegraphics[scale=1]{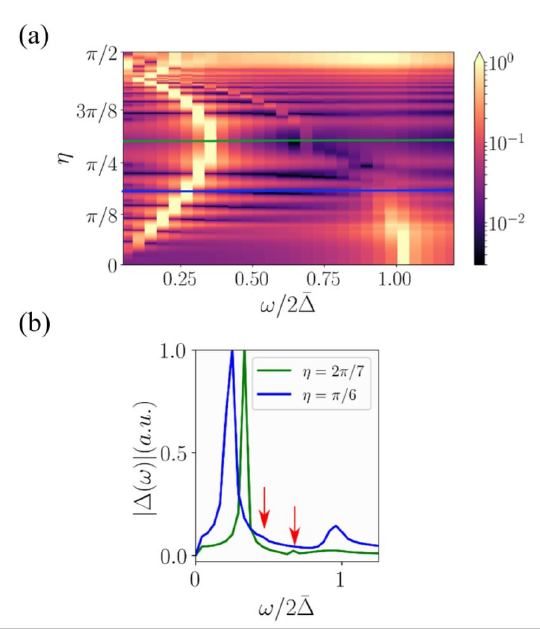} 
    \caption{ Numerical simulation of the oscillations following a state quench with strength $\delta$=0.2 and relative Fourier spectrum $|\Delta(\omega)|=|FT|\Delta(t)||$ (in arbitrary units) as a function of frequency $\omega$ (scaled to the asymptotic value of the gap $2\bar{\Delta}$) for a $s+id_{x^2 - y^2} $ superconductor in the $A^{2g}$-channel. (a) Collective spectra of the oscillation in the $A^{2g}$ channel, (b) Fourier transform of the oscillations corresponding to two line cuts of the plot in (a) at $\eta=\pi/6$ and $\eta=2\pi/7$ to show the additional weak peak associated with the quench dynamics, as in Ref.~\cite{lukas}. } 
    \label{fig:sid_A2g} 
\end{figure}

In our numerical investigations, the $s+id_{x^2-y^2}$ case is characterized by the fact that we are never able to disentangle the phase and amplitude sector by applying the quenches in the different channels listed in Tab.~\ref{tab:quenches}.\\
As one could expect by looking at the oscillations of the condensate triggered in the different channels, see Fig.~\ref{fig:sid_osc}, for the case of the s+id, no combination allows us to solely excite the phase or the amplitude sector. The two sectors, when triggered, are always simultaneously excited. In Figs.~\ref{fig:sup_sid},~\ref{fig:sid_A2g} we report, respectively, the plots corresponding to the excitation of the $A^{1g}$ and $A^{2g}$, which, both in the case of d+id' and p+ip', lead to well distinguishable features. In this scenario, there are no particularly distinguishable features. In Fig.~\ref{fig:sup_sid}(a), we display the collective spectra of excitation calculated by evolving in time the full equations of motion for the $A^{1g}$ oscillation; the spectrum is the same as the one obtained with the calculation up to linear order. In  Fig.~\ref{fig:sup_sid}(b), we show the time evolution of the gap, as it is possible to see the oscillations are persistent in time, due to the excitation being below the quasi-particle continuum.\\

In Fig.~\ref{fig:sid_A2g}(a), we plot the spectrum corresponding to the $A^{2g}$ oscillation in log-scale, in (b), we show the Fourier spectra for two line cuts of the 2d-plot at $\eta=2\pi/7$ and $\eta=\pi/6$.  The presence of an additional peak is shown; this peak is dynamically created and is associated with the quench dynamics as in Ref.\cite{lukas}.
\section{p+ip' -wave superconductor}\label{app:pip}
\begin{figure}[htpb!] 
    \centering 
\includegraphics[scale=.88]{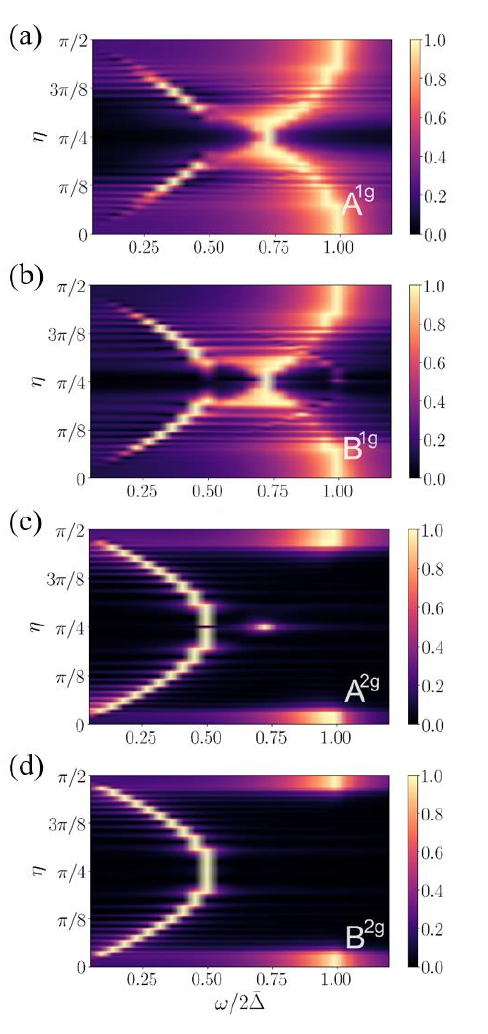} 
    \caption{ Numerical simulation of the Higgs oscillations following a state quench with strength $\delta$=0.2 and Fourier spectrum $|\Delta(\omega)|=|FT|\Delta(t)||$ (in arbitrary units), as a function of frequency $\omega$ (scaled to the asymptotic value of the gap $2\bar{\Delta}$) and the mixing angle $\eta$ for a p+ip' superconductor. From top to bottom: (a) $A^{1g}$-oscillations, (b) $A^{2g}$-oscillations, (c) $B^{1g}$-oscillations, (d) $B^{2g}$-oscillations.}  
    \label{fig:pip_quench} 
\end{figure}

\begin{figure}[htpb!] 
    \centering 
\includegraphics[scale=.9]{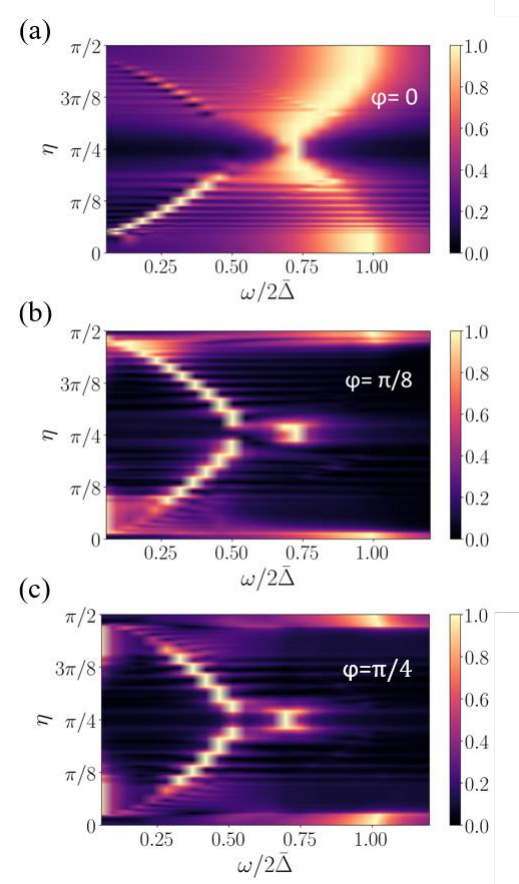} 
    \caption{ Numerical simulation of the condensate oscillations for a p+ip' order parameter following the application of a pump pulse at incident angles $\varphi=0, \varphi=\pi/8, \varphi=\pi/4$. Fourier spectrum $|\Delta(\omega)|=|FT|\Delta(t)||$ (in arbitrary units), as a function of frequency $\omega$ (scaled to the asymptotic value of the gap $2\bar{\Delta}$) and the mixing angle $\eta$. The pulse parameters are $\tau_p = 0.4$~ps, $|\mathbf{A}_p| = 7~10^{-8} \text{Js}~\text{C}^{-1} \text{m}^{-1} $ and $\hbar \omega = 4$~meV. For each $\eta$ value, the spectrum of the oscillations is normalized to the maximum value in the visualized frequency range.} 
    \label{fig:pip_pp} 
\end{figure}

Similarly to the d+id' case treated in the main text, we present here the results for the p+ip' in the spinless (effectively $S_z=0$) case,in two-dimensions. This particular kind of superconducting state can potentially be realized in ultra-cold atom experiments and in other engineered platforms~\cite{PRXQuantum.3.040324,PhysRevB.83.184520,PhysRevLett.100.096407} as it is interesting for its topological properties~\cite{PhysRevLett.126.237002,Alicea_2012}. In Fig.~\ref{fig:pip_quench} we show the results of the quench dynamics. In Fig.~\ref{fig:pip_pp} we show the results obtained by pumping the system in two relevant directions, $\varphi=0,\pi/8$. We compare the results obtained by the numerics in Table~\ref{tab:pip}. We note here that although the structure of the amplitude and phase sector is the same between the d+id' and p+ip' case, as it emerges from the Ginzburg Landau, the two order parameters are intrinsically very different: the p+ip' is associated with the ($E_u$) multidimensional irreducible representation (irrep) of the point group symmetry of the system characterized by odd parity basis functions and it has $C_2$ rotational symmetry. In order to quench the symmetry and excite even parity oscillations, we then had to use odd parity quench functions, see Table~\ref{tab:quenches}. By exciting the modes within our pump-probe scheme (Sec.~\ref{Sec:pp}) we are able to distinguish between the d+id' and the p+ip' case, while in the d+id' scenario at the pumping angles $\varphi=\pi/8$ and $\varphi=\pi/4$ correspond drastically different spectra, see Fig.~\ref{fig:did_pumpprobe}, for the p+ip'-case this is no more true, as $\pi/4$ does not represent a symmetry axis for the system, see Fig.~\ref{fig:pip_pp} The comparison between pump probe Fig.~\ref{fig:pip_pp} and group theory classification Fig.~\ref{fig:pip_quench} for the p+ip'-case is summarized in Table~\ref{tab:pip}.  

{\def\arraystretch{2}\tabcolsep=3 pt
\begin{table}[htpb!] 
    \centering
    \begin{tabular}{c|c|c}
    \toprule
        Gap & Condensate osc& Pump angle  \\\midrule
        
        $p_{x} + ip_{y}$ &  $A^{1g}_{p+ip'}$ & -\\
                &  $B^{1g}_{p+ip'}+A^{1g}_{p+ip'}$ & 0\\
                &  $A^{2g}_{p+ip'}+A^{1g}_{p+ip'}$ & $\pi/8,\pi/4$\\
                & $B^{2g}_{p+ip'}+A^{1g}_{p+ip'}$ & -\\
                \bottomrule
                
    \end{tabular}
    \caption{
    Comparison between pump-probe and group theory analysis for the p+ip' case. The second column lists all fundamental condensate oscillations. The third column lists the pumping angle able to induce the corresponding oscillation.\\}
    \label{tab:pip} 
\end{table}
}
 
\newpage
\null 
\newpage
\bibliography{bibliography.bib} 
\end{document}